%% file: main.tex
\documentclass[10pt, conference, letterpaper]{IEEEtran}

\usepackage{wrapfig,enumerate}
\usepackage{cite}
\usepackage[numbers]{natbib} 

\usepackage{algorithm}
\usepackage{xspace}

  \usepackage{graphicx}
\usepackage{epstopdf}

\usepackage{amsmath,amsthm,amssymb}
\usepackage{color,url}

\usepackage{comment}

\input{gmacros}

\usepackage[tight,footnotesize]{subfigure}

\def\E{\mathbb{E}}

\newcommand{\newcris}{}

\begin{document}

\title{Wireless Optimisation via Convex Bandits: \\ Unlicensed LTE/WiFi Coexistence}

\author{Cristina Cano$^1$ and Gergely Neu$^2$\\
\small
$^1$Universitat Oberta de Catalunya, Barcelona, Spain,\\
$^2$Universitat Pompeu Fabra, Barcelona, Spain.

}

\maketitle

\IEEEpeerreviewmaketitle
\input{abstract}

\input{introduction}

\input{seqbco}

\input{formulation}

\input{results}

\input{conclusions}

\bibliographystyle{IEEEtranN}
\bibliography{references}

\appendix
\input{appendix}

\end{document}

%% file: gmacros.tex
\newcommand{\real}{\mathbb{R}}

\newcommand{\EE}[1]{\mathbb{E}\left[#1\right]}

\newcommand{\abs}[1]{\left|#1\right|}
\newcommand{\norm}[1]{\left\|#1\right\|}

\newcommand{\ev}[1]{\left\{#1\right\}}
\newcommand{\pa}[1]{\left(#1\right)}
\newcommand{\bpa}[1]{\bigl(#1\bigr)}
\newcommand{\Bpa}[1]{\Bigl(#1\Bigr)}

\newcommand{\wt}{\widetilde}

\newcommand{\ra}{\rightarrow}

\usepackage{todonotes}
\definecolor{PalePurp}{rgb}{0.66,0.57,0.66}

\newcommand{\redd}[1]{{#1}}

\newcommand{\K}{\mathcal{K}}

\newcommand{\tg}{\wt{g}}
\newcommand{\tz}{\tilde{z}}
\newcommand{\wth}{\wt{\varphi}}

\newtheorem{assumption}{Assumption}
\newtheorem{theorem}{Theorem}
\newtheorem{corollary}{Corollary}

\newcommand{\ogdsemp}{\textsc{OGD-SeMP}\xspace}

%% file: abstract.tex
\begin{abstract}

Bandit Convex Optimisation (BCO) is a powerful framework for sequential decision-making \redd{in non-stationary and partially observable 
environments}. In a BCO problem, a decision-maker sequentially picks actions to minimize the cumulative cost associated with these 
decisions, all while receiving partial feedback about the state of the environment. 
This formulation is a very natural fit for wireless-network optimisation problems and has great application potential since: 
\emph{i)} \redd{instead of assuming full observability of the network state,} it only requires the metric to optimise as input, 
and \emph{ii)} it \redd{provides strong performance guarantees while making only minimal assumptions about the network dynamics}.
Despite these advantages, BCO has not yet been explored in the context of wireless-network optimisation. 
In this paper, we make the first steps to demonstrate the potential of BCO techniques by formulating an unlicensed LTE/WiFi fair 
coexistence use case in the framework, and providing experimental results in a simulated environment. On the algorithmic front, we 
propose a simple and natural sequential multi-point BCO algorithm amenable to wireless networking optimisation, and provide its theoretical 
analysis.
We expect the contributions of this paper to pave the way to further research on the application of online convex methods in the bandit setting.

\end{abstract}

%% file: introduction.tex
\section{Introduction}\label{sec:introduction}

Bandit Convex Optimisation (BCO) is structured as a repeated game \redd{between a decision-maker and its environment (often also called 
\emph{the adversary}): in each round, the learner picks a point in a convex set $\K$ and, simultaneously, the environment chooses a convex 
cost function that maps points in $\K$ to their corresponding cost. At the end of the round, the player incurs the cost 
associated with its action. A crucial feature of this setup is that \emph{the only} feedback that the player receives is its 
own cost: the adversary never reveals the entire cost function or even the gradient to the player, thus making for a very difficult 
learning problem for the player.
Were the gradients 
observed, the player could rely on the generic framework of Online Convex Optimisation (OCO)  and be able to 
efficiently optimise its costs by one of the many available standard algorithms for this setting \cite{hazan2016introduction}. Most 
literature on Bandit Convex Optimisation is concerned with accurate estimation of the gradients from the observed costs and reducing the 
learning problem to a more standard OCO setup.
BCO is a very active research area concerned with this problem, with the main focus on the design and rigorous theoretical performance 
analysis of algorithms.} However, practical application of these algorithms is still limited; there is very little known about their 
performance in practice for different application domains.

Alongside these developments, the wireless-networking community is also taking advantage of convex-optimisation 
techniques: the fact that convex objective functions can be optimised efficiently has motivated the formulation of many wireless network 
optimisation problems as convex ones. 
%
%
To mention a few, we can find convex formulation and application of convex optimisation methods in spectrum-sharing 
\cite{zhang2014resource}, cell-free massive 
MIMO systems \cite{nayebi2017precoding}, caching \cite{cui2017analysis}, D2D transmissions \cite{cheng2014optimal}, power allocation 
\cite{shen2014power} and wireless-powered networks \cite{ju2014throughput}, among many others.
\newcris{In some cases, the solution of these convex-optimisation problems is explicit, while in others an iterative optimisation algorithm is needed.
In either case implementation in practice is usually feasible but requires knowledge of the convex function.
This entails inferring network conditions and complicates handling network dynamics (i.e., nodes entering/leaving the network and varying 
channel conditions, among others).} 

\newcris{
\redd{In this article, we champion the BCO framework as a natural remedy for these practical challenges: as BCO readily addresses the issue 
of partial feedback from the environment, it obviates the need to infer network conditions.
As an example consider 
traditional problem formulations concerning proportional fairness in WiFi networks } (such as those in 
\cite{paul2014rigorous,valls2014proportional,checco2011proportional,liew2008proportional}).
In these examples implementing the optimal solution requires retrieving network measurements and computing the channel access parameters that equalise the channel airtimes of the WiFi stations. 
Using BCO instead, the per-station throughput would be the only input required to the BCO algorithm, which is a readily available metric at the access point. 
Additionally, network dynamics such as a varying number of WiFi stations and changes in the packet transmission duration are handled by BCO 
intrinsically as it \redd{handles non-stationary sequences of cost functions by design}.
}
Despite these advantages, as we mentioned, BCO has not yet been explored to 
solve wireless network optimisation problems. The reason for this is arguably the current lack of practical evaluations.

In this article we derive a new BCO algorithm suitable for wireless network optimisation by extending multi-point BCO methods to a 
sequential setting. We provide a theoretical analysis of this approach and evaluate its performance in practice using a topical use case. 
We consider the case of fair coexistence between unlicensed LTE and WiFi, which has attracted considerable attention in the last few years as fair coexistence among 
these networks is crucial given the heterogeneity of their channel accesses \cite{cano2016using,al20155g}.
Our results in this use case show that the performance of our proposed BCO approach is suitable for practical application.
We believe the contributions of this work will pave the way to further research on practical wireless optimisation algorithms based on 
online convex optimisation methods under partial information.

The remainder of this article is organised as follows. 
In Section~\ref{sec:seq_bco} we present our proposed multi-point sequential bandit approach as well as its theoretical analysis.  
Then, we formulate the unlicensed LTE/WiFi BCO problem in Section~\ref{sec:formulation}.
We show the performance results in Section~\ref{sec:results} and conclude the article in Section~\ref{sec:conclusions} with some final remarks.

%% file: seqbco.tex
\section{Bandit Convex Optimisation}\label{sec:seq_bco}
In a Bandit Convex Optimisation (BCO) problem, the following steps of interaction are repeated between a player and its environment (also 
referred to as \emph{the adversary}) for a 
number of rounds $t = 1,2,\dots,T$:
\begin{itemize}
 \item The player chooses a point $x_t \in \mathcal{K}\subseteq \real^d$.
 \item The environment chooses a loss function $f_t \in \mathcal{F} \subseteq \real^\mathcal{K}$.
 \item The player observes $f_t(x_t)$.
\end{itemize}
We assume that $\mathcal{K}$ is a convex subset of a $d$-dimensional Euclidean space and all functions in $\mathcal{F}$ are convex. In this 
paper, we will consider the simplest case $d=1$, which is sufficient for our application and greatly simplifies presentation. We are 
interested in constructing algorithms for the player that guarantee that the cumulative sum of the incurred losses $f_1(x_1), f_2(x_2),...$ 
is as small as possible. A common way of measuring the performance of learning algorithms in this setting is by means of the 
\emph{cumulative regret} (or, in short, \emph{regret}), which is defined as
\begin{equation}
 R_T = \sum_{t=1}^{T} f_t(x_t) - \min_{x \in \mathcal{K}} \sum_{t=1}^{T} f_t(x).
\end{equation}
The goal is to construct a learning algorithm that guarantees that the regret grows sublinearly, that is, that the 
average regret per round vanishes as $T$ grows large: $R_T/T \ra 0$ as $T\ra \infty$. Intuitively, sublinear regret means that the 
algorithm learns to perform as well as the best fixed point $x^*\in\K$ chosen in full knowledge of the sequence of cost functions.

The main difficulty that characterizes the BCO setting is that the player only observes the function values $f_t(x_t)$ in each round, and 
never observes the full cost functions $f_t$ played by the adversary, or even the (sub-)gradients of $f_t$ evaluated at $x_t$ (denoted as 
$\nabla f_t(x_t)$).
%
A large fraction of all work in BCO is concerned with developing methods for estimating these gradients and using the resulting estimates in 
conjunction with algorithms designed for Online Convex Optimisation (OCO, \citep{hazan2016introduction}). This approach was pioneered by 
\citet{FKM05,Kle05} who proposed a delicate randomized estimation scheme, and combined the estimated gradients with the Online Gradient 
Descent (OGD) algorithm of \citet{Zin03}. Since these works, several, even more intricate schemes were designed along the same 
lines---we refer to \citet{HPGS16} for a review of this line of work. Most relevant of these variants to our approach is the multi-point 
gradient estimation technique proposed by 
\citet{agarwal2010optimal}, who crucially assume that in each round, the learner can query $f_t$ at two different points. Based on 
this assumption, \citet{agarwal2010optimal} are able to construct an algorithm with a regret bound of $O(\sqrt{T})$, improving over 
the $O(T^{3/4})$ bounds proved by \citet{FKM05}. 

One concern about the two algorithms described above is that none of them are well-suited for practical implementation. First, we observed that the 
variance of the single-point estimators proposed by \citet{FKM05} is prohibitively large and the tiny step sizes required to offset this 
variance lead to impractical convergence speeds, even when optimising a fixed function. Second, the assumption of 
\citet{agarwal2010optimal} that one can query the objective functions multiple times per round is unrealistic in many settings.
Below, we propose a simple variant of the multi-point gradient estimation method that does not suffer from any of these limitations.

\subsection{Online Gradient Descent with Sequential Multi-Point Gradient Estimates}
We now describe a new BCO algorithm that uses a multi-point gradient estimation method that does not require the restrictive assumption 
made by \citet{agarwal2010optimal}. Unlike previous multi-point estimates, we combine queries from \emph{two consecutive rounds} 
to form our gradient estimate. Intuitively, this method will produce reliable gradient estimates as long as the sequence of loss 
functions changes in a limited fashion. \redd{Perhaps more surprisingly, we show below that the estimates produced by our scheme are 
reliable enough to guarantee sublinear regret without any significant assumption on the sequence of the loss functions. 
Specifically, we will provide a general analysis that bounds the regret in terms of the \emph{total deviation} of the loss functions, which 
allows us to recover the guarantees  of \citet{FKM05,Kle05} in the worst case, but also leads to improved guarantees in various cases of 
practical interest. 
} 

Throughout the section, we make the following mild assumption on the cost functions:
\begin{assumption}\label{as:lipschitz}
 For all $t=1,2,\dots,T$, the cost functions are Lipschitz-continuous: for all $x,y\in\K$, they satisfy
 \[
  \abs{f_t(x) - f_t(y)} \le G \norm{x-y}.
 \]
 Furthermore, all loss functions are bounded in $[0,C]$.
\end{assumption}

%
%

Formally, we will consider a one-dimensional decision set $\K = [A,B]$ with $D = B - A$ and define $\K_\alpha = [A+\alpha, 
B-\alpha]$ for any $\alpha\le D/2$. We also define $\Pi_{[a,b]}(x) = \max\ev{\min\ev{b,x},a}$ as the projection of $x$ to the 
nonempty interval $[a,b]$. Our algorithm will maintain a sequence of auxiliary points $y_1,y_2,\dots$ and query the loss 
functions around said points to compute gradient estimates $\tg_1,\tg_2,\dots$, which will be used to update $y_k$ as
\[
 y_{k+1} = \Pi_k\pa{y_{k} - \eta_k \tg_k}
\]
for some appropriately defined projection operator $\Pi_k$. We will refer to this algorithm as \textbf Online \textbf Gradient \textbf 
Descent with \textbf{Se}quential \textbf{M}ulti-\textbf{P}oint Gradient Estimates, or, in short, \ogdsemp. The precise algorithm is 
presented as Algorithm~\ref{alg:ogdsemp}.

\begin{algorithm}
\caption{\ogdsemp}\label{alg:ogdsemp}
 \textbf{Input parameters:} Non-increasing sequences $\pa{\delta_k}, \pa{\eta_k}$.
 \\
 \textbf{Initialization:} Choose arbitrary $y_0 \in \K_{\delta_0}$.
 \\
 \textbf{For $k=0,1,\dots,$ repeat:}
 \begin{enumerate}
  \item Draw $\varepsilon_k$ uniformly from $\ev{-1,1}$.
  \item Let $t=2k+1$ and play 
  \[
    x_t = y_k + \varepsilon_k \delta_k.
  \]
  \item Set $g^+_k = f_t(x_t)$.
  \item Play 
  \[
    x_{t+1} = y_k - \varepsilon_k \delta_k.
  \]
  \item Set $g^-_k = f_{t+1}(x_{t+1})$.
  \item Compute gradient estimate
  \[
   \tg_k = \frac{g^+_k - g^-_k}{2\varepsilon_k \delta_k}
  \]
  \item Update
  \[
   y_{k+1} = \Pi_{k}\pa{y_k - \eta_k \tg_k},
  \]
  where $\Pi_k$ is the projection operator onto $\K_{\delta_k}$
 \end{enumerate}
\end{algorithm}

\subsection{Analysis}
We now provide a theoretical analysis for \ogdsemp. For simplicity of exposition, we focus on 
time-invariant parameters $\eta$ and $\delta$, noting that proving similar results for non-increasing sequences $\pa{\eta_k}$ and 
$\pa{\delta_k}$ is also possible at the expense of slightly more involved derivations. Without loss of generality, we will assume that 
$T$ is even. 

For the sake of analysis, we will consider a stronger notion of regret that we call \emph{interval regret}, defined as
\[
 R_{[s,r]} = \sum_{t=s}^{r} f_t(x_t) - \min_{x \in \mathcal{K}} \sum_{t=s}^{r} f_t(x)
\]
over any interval $[s,r]$. The merit of this regret notion is that it measures the performance of the algorithm against that of the best 
fixed decision \emph{for every interval}, and not just the best single decision that is optimal for the entire interval $[1,T]$. Our main 
result below will rely on the definition of the \emph{instantenous deviation} of the functions defined as
\[
 \alpha_k = \sup_{x\in\K} \abs{f_{2k+1}(x) - f_{2k+2}(x)}
\]
for all $k=0,1,\dots,T/2$. We note that $\alpha_k$ is always trivially bounded by $2C$. The \emph{total deviation} of any subsequence of 
loss functions $f_s,f_s+1,\dots,f_r$ will be defined as
\[
 L_{[s,r]} = \sum_{k=(s-1)/2}^{(r-1)/2} \alpha_k^2.
\]
The theorem below bounds the interval regret of \ogdsemp for any nonempty interval $[s,r]\subseteq[1,T]$ in terms of $L_{[s,r]}$.
\begin{theorem}\label{thm:main}
 Let $\eta_k = \eta > 0$ and $\delta_k = \delta \in (0,D/2)$ for all $k$ and suppose that the loss functions satisfy 
Assumption~\ref{as:lipschitz}. Let $s$ and $r$ be two odd time steps satisfying $s<r$, let $s' = (s-1)/2$, $r' = (r-1)/2$, and $\Delta = 
r-s$. Then, the expected interval regret of \ogdsemp over $[s,r]$ can be bounded as
\[
 \EE{R_{[s,r]}} \le \frac{2 D^2}{\eta} + \eta G^2 \Delta + \frac{\eta L_{[s,r]}}{4\delta^2} + 4\delta G \Delta.
\]
\end{theorem}
The proof is given in Appendix~\ref{app:proof}. All of our theoretical results in the paper will be consequences of Theorem~\ref{thm:main}.
As a first demonstration of the usefulness of this theorem, we first state a general regret bound that holds without 
further assumptions on the loss functions:
\begin{corollary}\label{cor:general}
 Suppose that the loss functions satisfy Assumption~\ref{as:lipschitz}. Then, setting $\eta = \frac{G}{DT^{3/4}}$ and $\delta = C 
T^{-1/4}$, the expected regret of \ogdsemp satisfies
\[
 \EE{R_T} \le \pa{2GD + 4GC + \frac{G}{4D}} T^{3/4} + GD T^{1/4}.
\]
\end{corollary}
The corollary follows easily from setting $s=1$ and $r=T+1$ in Theorem~\ref{thm:main}, and observing that $L_{[1,T+1]}\le 
C^2T$. This performance guarantee essentially matches the best known results in the worst case where no further 
assumptions are made about the loss functions \citep{FKM05,Kle05}. Perhaps surprisingly, this shows that our sequential multi-point 
estimates work well even when the consecutive loss functions can change arbitrarily. Below, we discuss some special cases in which our 
bounds improve over this result.

\subsubsection{Infrequently changing losses}
We first study the case where the loss functions are changing infrequently in the following sense:
\begin{assumption}\label{as:piecewise}
For all $t=1,2,\dots,T$, the cost function can be written as
\[
 f_t(x) = \varphi_{\tau(t)}(x),
\]
for all $x\in\K$, where $\varphi_1,\varphi_2,\dots,\varphi_N$ are convex functions defined over $\K$, and $\tau:\ev{1,2,\dots,T}\rightarrow 
\ev{1,2,\dots,N}$ is a nondecreasing function.
\end{assumption}
In plain words, Assumption~\ref{as:piecewise} stipulates that the cost function changes at most $N$ times during $T$ rounds. The following 
performance guarantee is a useful improvement over Corollary~\ref{cor:general} when $N\ll T$:
\begin{corollary}\label{cor:dynamic_total}
 Suppose that the loss functions satisfy Assumptions~\ref{as:lipschitz} and~\ref{as:piecewise}. Then, setting $\eta = \frac{G}{D\sqrt{T}}$ 
and $\delta = \frac{C \ln T}{T}$, the expected regret of \ogdsemp satisfies
\[
 \EE{R_T} \le (N/2 + 3) GD \sqrt{T} + 4CG \ln T.
\]
\end{corollary}
 The proof follows from applying Theorem~\ref{thm:main} with $[s,r]=[1,T+1]$ and observing that $L_{[1,T+1]}\le 2NC^2$, which yields
 \[
  \frac{\eta L_{[s,r]}}{4\delta^2} \le \frac{\eta 2NC^2}{4\delta^2} \le \frac{N\sqrt{T}}{2\ln^2 T}.
 \]
The bound scales linearly with the number of switches of the objective function, which is expected given that our two-point gradient 
estimators will fail to estimate the gradient every time that the loss function changes, thus incurring some extra cost upon every switch. 
Nevertheless, the bound remains meaningful in the domain where $N\ll T$, that is, when the loss function changes rarely. 

Besides the above result, Theorem~\ref{thm:main} also allows us to bound the regret within every interval where the loss function remains 
unchanged:
\begin{corollary}\label{cor:dynamic}
 Suppose that the loss functions satisfy Assumptions~\ref{as:lipschitz} and~\ref{as:piecewise}. Let us fix any interval $[s,r]$ such that 
$\tau(s) = \tau(r)$ holds and let $\Delta = r-s$. Then, setting $\eta = \frac{G}{D\sqrt{T}}$ 
and $\delta = \frac{C \ln T}{T}$, the expected interval regret of \ogdsemp can be bounded as
\[
 \EE{R_{[s,r]}} \le 3GD\sqrt{T}  + 4CG\ln T + 2C.
\]
\end{corollary}
The proof follows from applying Theorem~\ref{thm:main} to the interval $[s,r]$, with the slight technical obstacle that $s$ and $r$ may not 
be both odd. This issue is handled by considering the closest odd $s$ and $r$ within the original interval and bounding the regret in the 
offending rounds by $C$.
Notably, this result is much stronger than Corollary~\ref{cor:dynamic_total}, as it compares the performance of the algorithm to the best 
fixed decision within each period where the losses are unchanged, whereas Corollary~\ref{cor:dynamic_total} uses the best fixed decision 
computed for the entire period $[1,T]$. 

The intuitive lesson from these corollaries is that if the loss function remains unchanged for long periods of time, then one can safely 
choose a relatively large learning rate of order $T^{-1/2}$ and an arbitrarily small exploration parameter $\delta$. Notably, the 
$O\bpa{\sqrt{T}}$ rate of the regret bound matches the one of \citet{agarwal2010optimal} obtained for gradient estimators that can query 
the loss function in 2 points \emph{within each round}. 

\subsubsection{Analysis under slowly changing losses}
We now let the functions $\pa{f_t}$ change continuously, with the restriction that they should change slowly in the following 
sense:
\begin{assumption}\label{as:slow}
 For all $t=1,2,\dots,T$, the loss functions satisfy
\[
 \sup_{x\in\K} \abs{f_t(x) - f_{t+1}(x)}\le \alpha.
\]
\end{assumption}
Under this assumption, we prove the following performance guarantee for \ogdsemp:
\begin{corollary}\label{thm:slow}
 Let $\eta_k = \eta > 0$ and $\delta_k = \delta \in (0,D/2)$ for all $k$ and  
suppose that the loss functions satisfy Assumptions~\ref{as:lipschitz} and~\ref{as:slow}. Then, setting $\eta = \frac{G}{DT^{3/4}}$ and 
$\delta = \alpha T^{-1/4}$, the expected regret of \ogdsemp satisfies
\[
 \EE{R_T} \le \pa{2GD + 4G\alpha + \frac{G}{4D}} T^{3/4} + GD T^{1/4}.
\]
\end{corollary}
This is a simple corollary of Theorem~\ref{thm:main} when $\alpha_k \le \alpha$ for some fixed $\alpha$ for all $k$.
The advantage of this result over Corollary~\ref{cor:general} is that it improves the undesirable $GC$ factor to $G\alpha$ when the change 
rate of the losses is small.\footnote{Indeed, in the worst case, $C$ can be as large as $GD$, which yields to a suboptimal dependence on 
$G$ in the final bound.} As in the previous case, this improvement is achieved by setting a smaller exploration parameter than the one 
suggested by Corollary~\ref{cor:general}. This suggests the intuitive lesson that more ``regular'' loss sequences allow 
to get away with less exploration.
%
%
%
%
%

%% file: formulation.tex
\section{Coexistence of Unlicensed LTE and WiFi}\label{sec:formulation}

\newcris{
We describe in this section the application example of coexisting unlicensed LTE and WiFi networks in the 
same frequency band. \redd{While this problem is very well-studied in stationary environments where the WiFi network parameters are fixed 
and known, 
we consider here a more realistic and challenging setting where the environment may be partially observable and changing over time in an 
arbitrary fashion. As the stationary solution can be formulated in terms of a convex optimisation problem, the online problem we consider is 
naturally cast as an instance of Online Convex Optimisation. Moreover, as the practical limitations posed by partial observability 
are naturally modeled by the bandit framework, we will rely on the formalism of Bandit Convex Optimisation. The subsections below 
specify the details of our setting, the convex formulation of the stationary problem, and the online optimisation problem we aim to 
tackle.}
}

\newcris{\subsection{Scope}}

\newcris{
The use of unlicensed spectrum by mobile network operators, particularly in the 5 GHz band, is attracting considerable attention as it can 
assist \redd{in} satisfying increasing traffic demands.
However, the risks of employing legacy LTE in unlicensed bands without proper access control that ensures fair coexistence to WiFi networks 
\redd{have been pointed out several times in the literature (see, e.g., \cite{cavalcante2013performance,ltevswifi-01,lteu-exps})}.  
To enable this fair coexistence there are two main approaches under consideration at present.  
Namely, \emph{Listen Before Talk} (LBT) \cite{3gppstudy} and \emph{Carrier Sensing and Adaptive Transmission} (CSAT) \cite{sadek2015extending}.   
LBT uses carrier sensing while CSAT schedules transmissions according to a specified duty-cycle, oblivious to the channel status.
Several works have studied the ability of these approaches to provide different notions of fairness to WiFi 
\cite{ning2012unlicensed,hajmohammad2013unlicensed,liu2014small,ccano-icc,guan4cu}.
The proportional fair optimal channel time allocation has been derived by \citet{ccano-icc-2016,ccano-ton}.
We build upon these works and adapt the optimisation framework of \citet{ccano-ton} to the BCO algorithm presented in the previous section, 
\redd{demonstrating the practical usefulness of online optimisation in this setting.}
}

\newcris{\subsection{Proportional Fair Solution}}

\newcris{
Consider an LTE base station and a set of random access WiFi devices coexisting in the same frequency band. The LTE base station, following 
either LBT or CSAT, is active during $T_{\rm on}$ and remains silent during $\bar{T}_{{\rm off}}$, with $\bar{T}_{{\rm off}}:=\E[T_{\rm 
off}]$, $T_{{\rm off},k}$, $k=1,2,\dots$ denote the duration of the $k$-th off intervals and are i.i.d random variables. To simplify 
presentation we consider here the CSAT approach in which the LTE network transmits oblivious to the channel state (i.e., whether there is an 
ongoing transmission by WiFi). Generalisation to LBT is immediate following the formulation in \cite{ccano-ton}. We assume perfect channel conditions, no hidden terminals, saturation and perfect inter-technology detection.\footnote{For a 
discussion on these assumptions and on how they can be relaxed refer to \cite{ccano-ton}.} It has been shown in \cite{ccano-ton} that under 
these assumptions the throughput of WiFi station $j = 1,...,n$ can be computed as follows:
}
\begin{align}\label{eq:s_wifi}
s_{{\rm wifi},j} &= s_j \frac{\bar{T}_{{\rm off}} - c_1}{T_{\rm on}+\bar{T}_{{\rm off}}}, 
\end{align}
\normalsize
\newcris{where $s_j$ is the usual throughput of a WiFi station when no LTE network is present (as in~\cite{checco2011proportional}), which 
is now scaled considering the LTE base station on-off transmission activity.  
Considering CSAT, $c_1$ in Eq. \ref{eq:s_wifi} captures the average airtime lost by WiFi due to a partial collision with an LTE transmission.
This airtime can be approximated as $c_1:=\frac{T_{\rm fra}}{2}p_{{\rm txA}}$, with $p_{{\rm txA}}$ the probability of an LTE transmission colliding with WiFi and $T_{\rm fra}$ the duration of a WiFi frame transmission.
}

\newcris{Similarly, in the LTE side, the throughput is given as follows \cite{ccano-ton}:}
\begin{align}
s_{{\rm LTE}}&= r\frac{T_{\rm on}-c_2}{T_{\rm on}+\bar{T}_{{\rm off}}}, 
\end{align}
\normalsize
\newcris{with $r$ the LTE data rate, and $c_2$ the average airtime lost by LTE due to collisions with WiFi at the start of an LTE on period.
Since LTE transmits frames (of duration $\gamma$) back-to-back, $c_2$ captures the duration of the frames experiencing collision instead of the full on period: $c_2:=\lceil\frac{T_{\rm fra}}{2\gamma}\rceil \gamma p_{{\rm txA}}$.}

Then, letting $z:=\bar{T}_{{\rm off}} - c_1$, $\tilde{z}:=\log z$, $\tilde{s}_{{\rm wifi},j}:=\log {s}_{{\rm wifi},j}$ and $\tilde{s}_{{\rm 
LTE}}:=\log {s}_{{\rm LTE}}$, ${s}_{{\rm wifi},j}$ and ${s}_{{\rm LTE}}$ are transformed into convex functions in $\tilde{z}$: 
$\tilde{s}_{{\rm wifi},j} = \log s_j +\tilde{z}-\log(T_{\rm on}+c_1+e^{\tilde{z}})$, and $\tilde{s}_{{\rm LTE}}=\log (r(T_{\rm on}-c_2)) 
-\log(T_{\rm on}+c_1+e^{\tilde{z}})$. 
\newcris{Taking a proportional fair approach, the aim of \cite{ccano-ton} was to find ${\tilde{z}}^*$ that 
minimises the following convex function:}
\begin{align}\label{eq:f}
f{(\tz)} := - \tilde{s}_{{\rm LTE}}{(\tz)} - \sum_{j=1}^{n} \tilde{s}_{{\rm wifi},j}{(\tz)}.
\end{align}
\normalsize
\newcris{
\redd{Given explicit knowledge of the WiFi network parameters,} the convex optimisation problem in Eq. \ref{eq:f} can be solved explicitly 
and provides an effective method to compute the value of $\bar{T}_{{\rm off}}$ the LTE base station should set to provide proportional 
fairness to WiFi. \redd{The section below addresses the case where the problem parameters are unknown and good solutions have to be learned 
online from partial information.}
}

\newcris{\subsection{Formulation as a BCO Problem}}

\newcris{We proceed to formulate this problem in a BCO framework with the goal to explore applicability of the online algorithm defined in Section \ref{sec:seq_bco}.
Consider a repeated game of $T$ rounds. 
In each round $t =1,2,\dots,T$ we consider that the WiFi network is formed by some number of WiFi stations using some channel access probabilities (depending on the level of contention), modulation and coding rates (depending on varying channel conditions) and packet size. These parameters affect the WiFi throughput, the transmission duration of a WiFi packet and the collision probability with LTE. Thus, in each round the WiFi network selects Eq. \ref{eq:f} with some $n$, $s_j$, $T_{\rm fra}$ and $p_{{\rm txA}}$. We denote this resulting function $f_t$. Then, in each round $t$:}
\newcris{
\begin{itemize}
 \item The LTE base station chooses a value for $\bar{T}_{{\rm off}}$ and computes $\tilde{z}_t \in \mathcal{K}$.
 \item The WiFi network independently selects $f_t \in \mathcal{F}$.
 \item The LTE base station observes $f_t(\tilde{z}_t)$.
\end{itemize}
}
The decision set $\mathcal{K}$ is convex, all functions in $\mathcal{F}$ have gradient:
\begin{align}
 g_t = (n + 1) \frac{e^{\tilde{z}_t}}{T_{\rm on}+c_1+e^{\tilde{z}_t}} - n,
\end{align}
\normalsize
and are convex in $\tilde{z}_t$. Thus, all conditions are in place for applying the algorithm presented in Section 
\ref{sec:seq_bco}. 
\newcris{Note that in addition to the cost function $f_t(\tilde{z}_t)$ the BCO algorithm requires $c_1$ to compute $\tilde{z}_t$ for this application example.
This can be done as described in \cite{ccano-ton}.
}

%% file: results.tex

\begin{figure*}[hhht!]  
\centering
\subfigure[$\omega = 0.01$.]{\includegraphics[width=0.65\columnwidth]{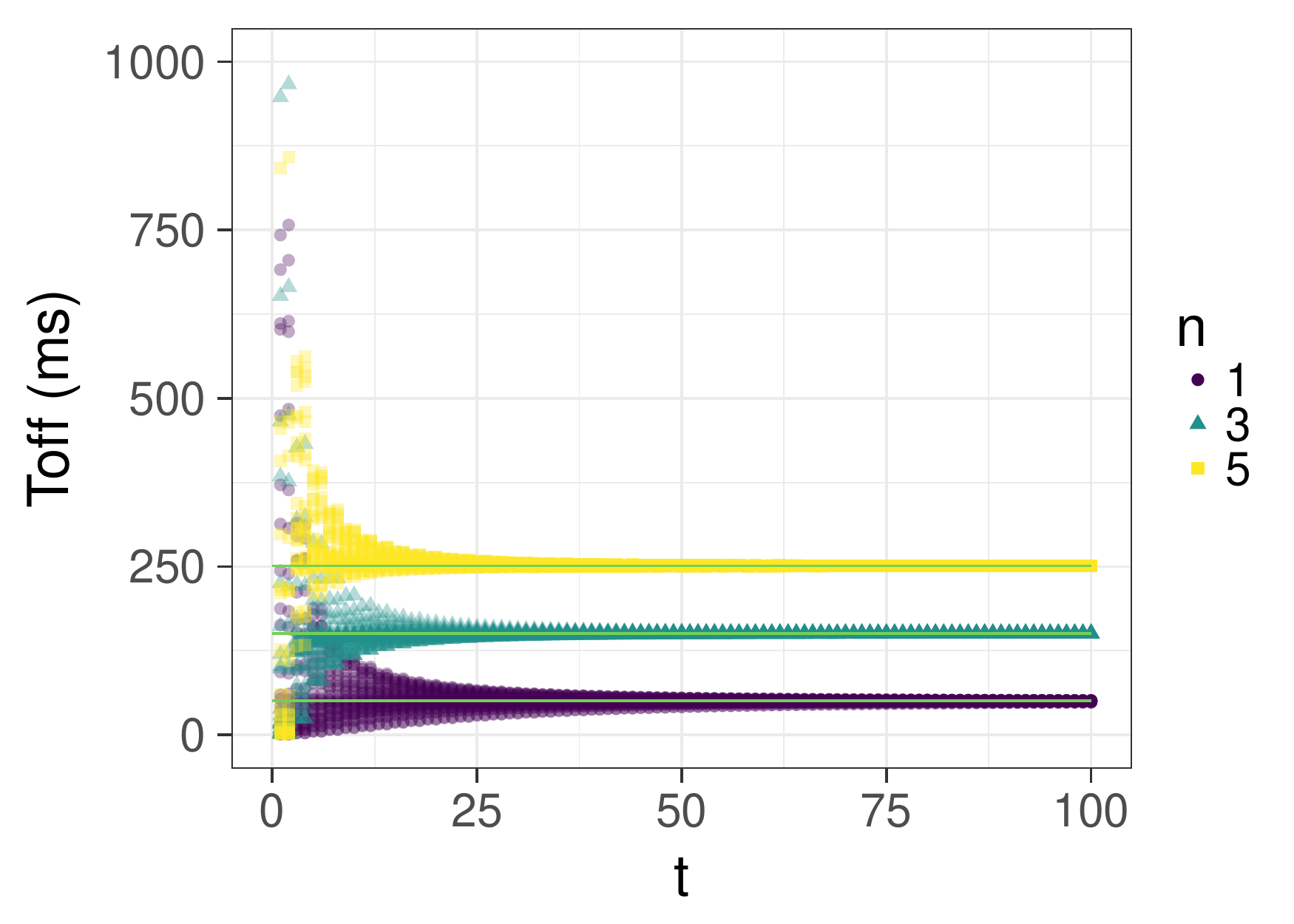}}
\subfigure[$\omega = 0.01$.]{\includegraphics[width=0.65\columnwidth]{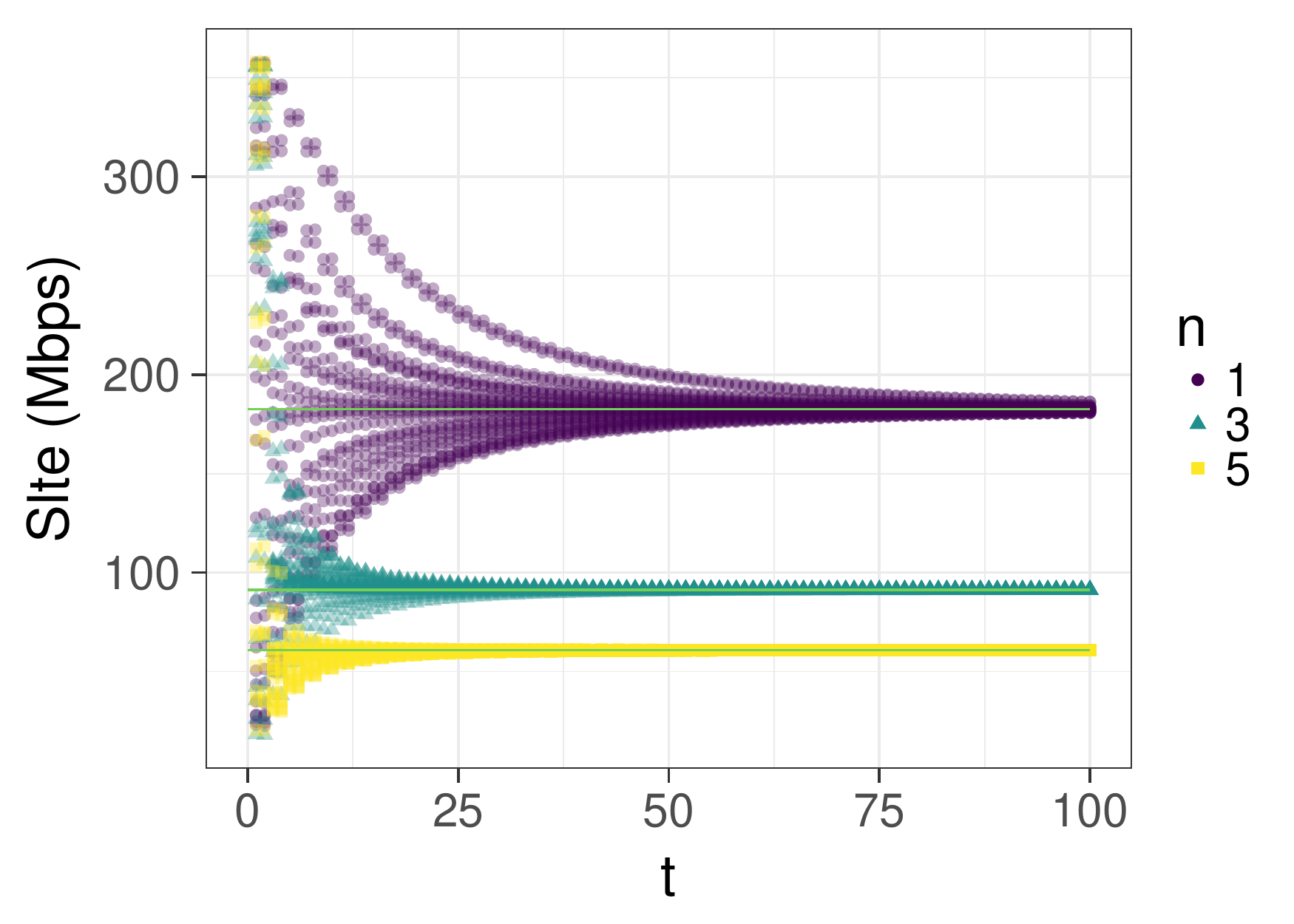}}
\subfigure[$\omega = 0.01$.]{\includegraphics[width=0.65\columnwidth]{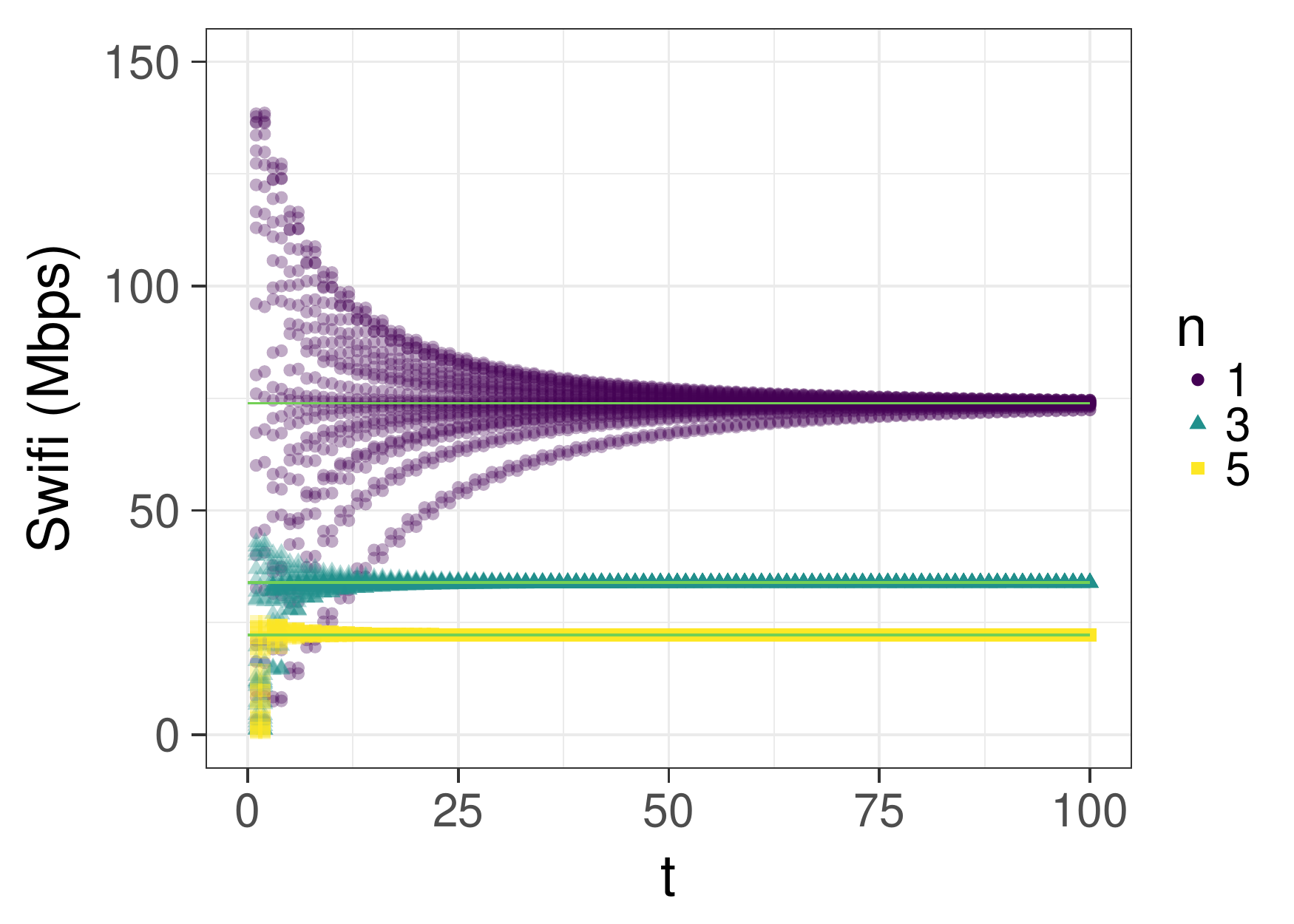}}\\
\subfigure[$\omega = 0.1$.]{\includegraphics[width=0.65\columnwidth]{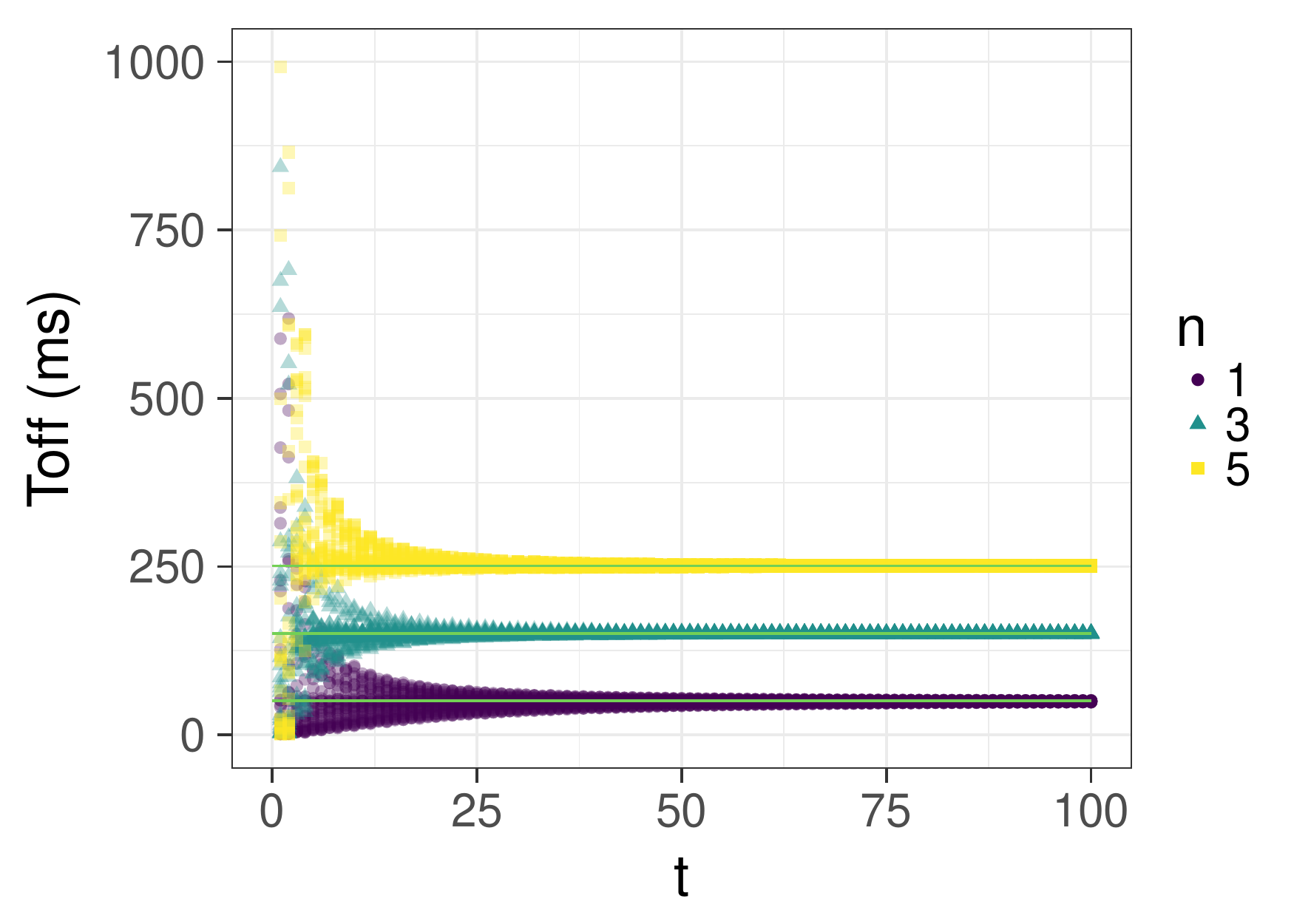}}
\subfigure[$\omega = 0.1$.]{\includegraphics[width=0.65\columnwidth]{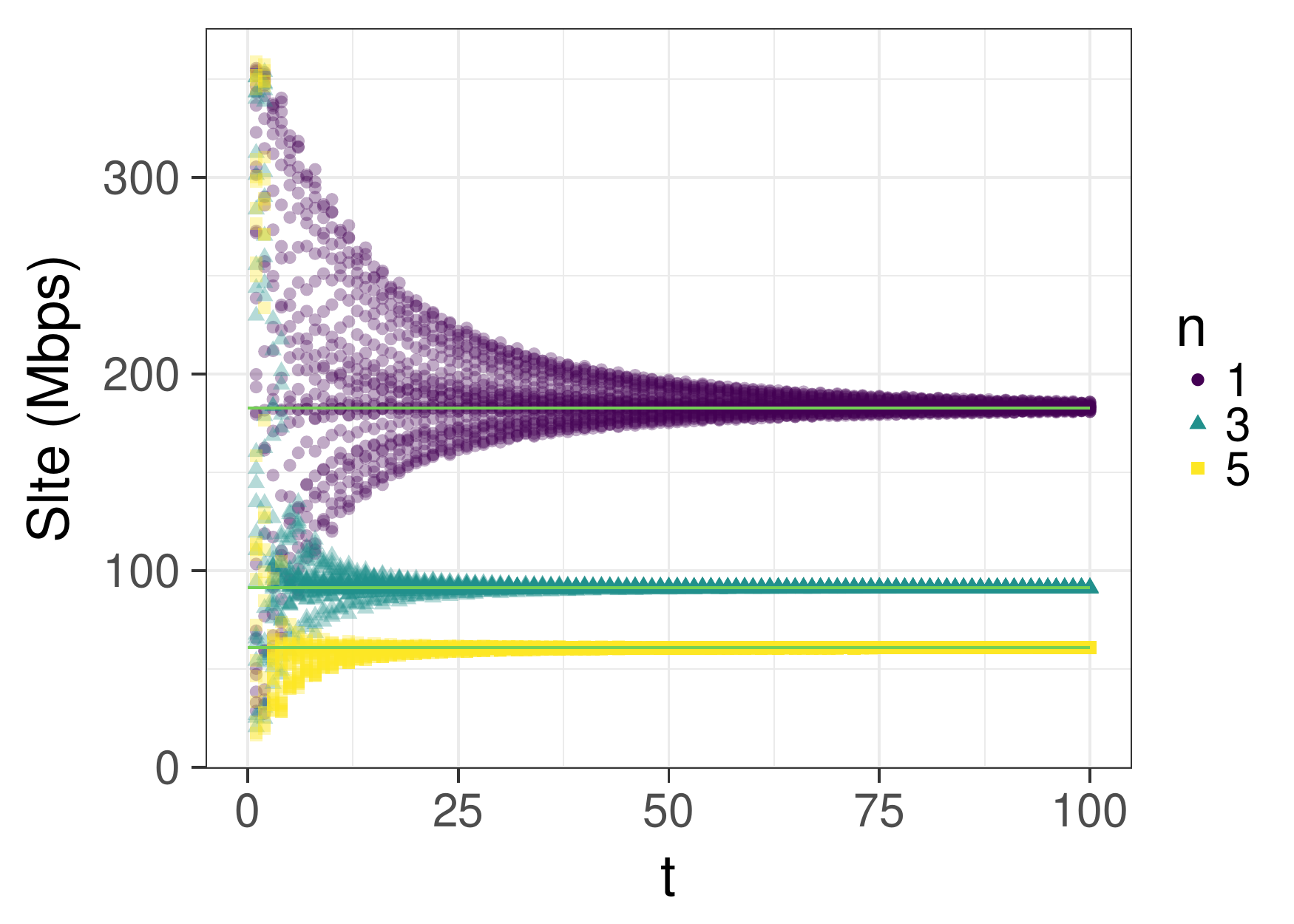}}
\subfigure[$\omega = 0.1$.]{\includegraphics[width=0.65\columnwidth]{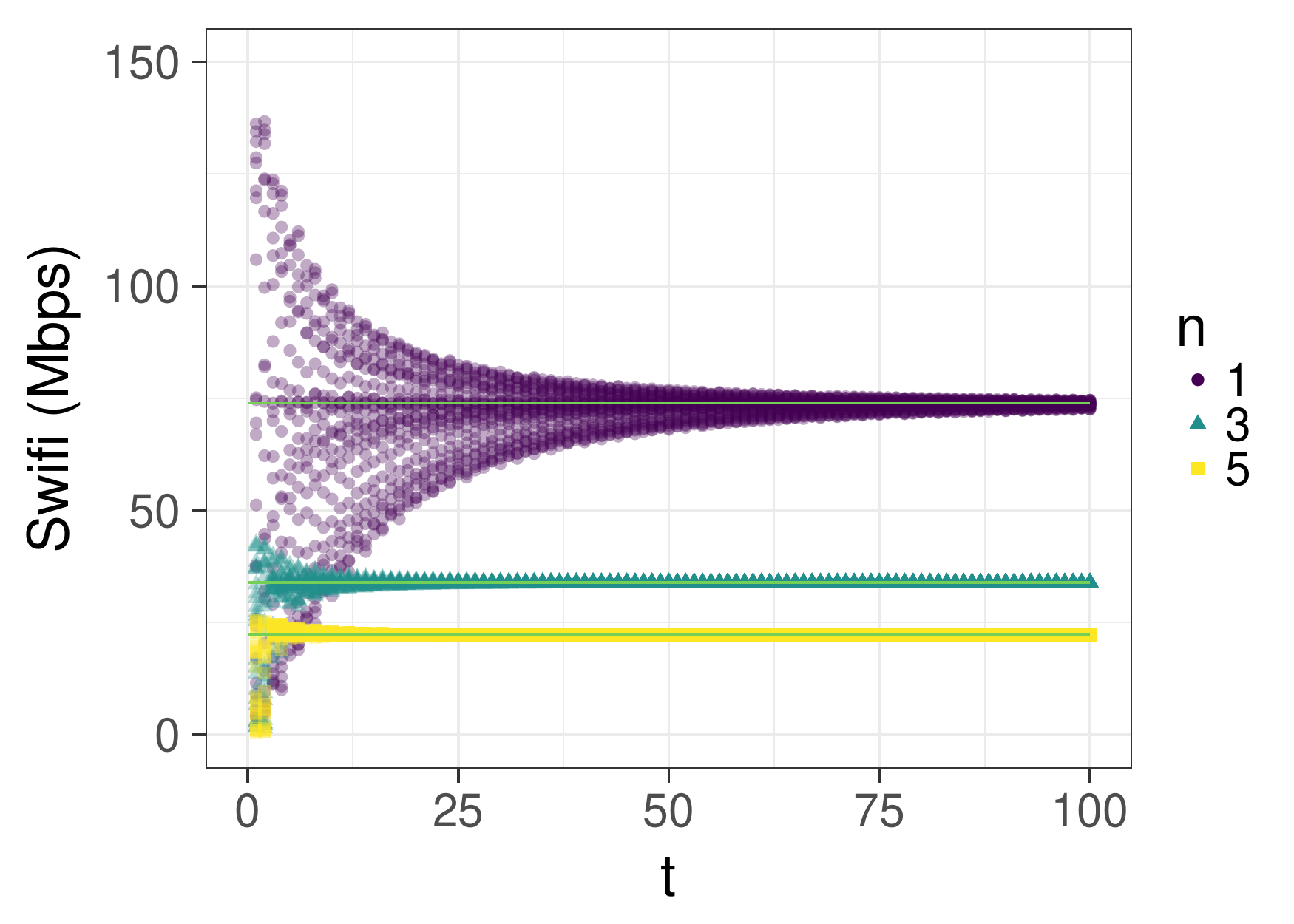}}\\
\subfigure[$\omega = 1$.]{\includegraphics[width=0.65\columnwidth]{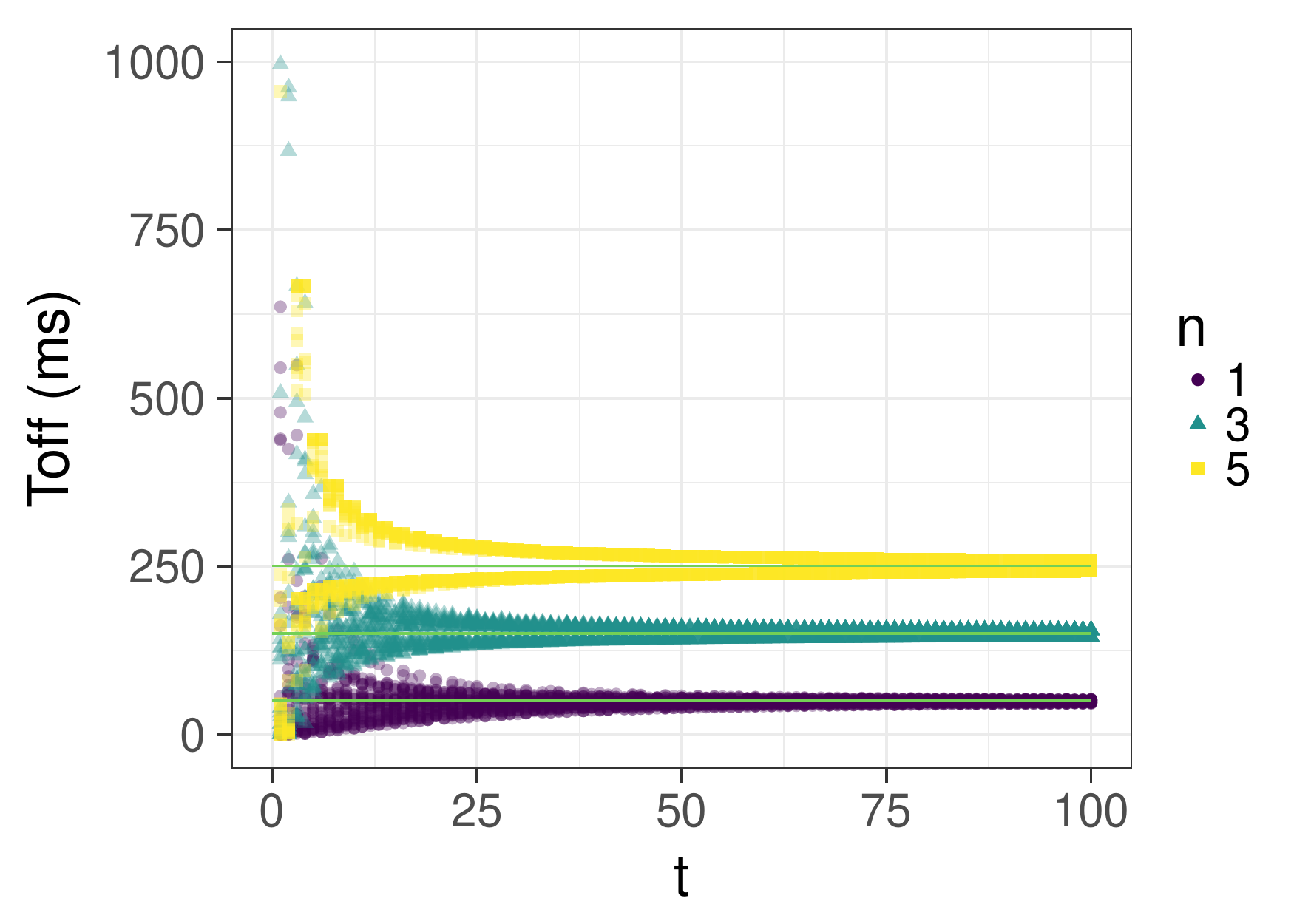}}
\subfigure[$\omega = 1$.]{\includegraphics[width=0.65\columnwidth]{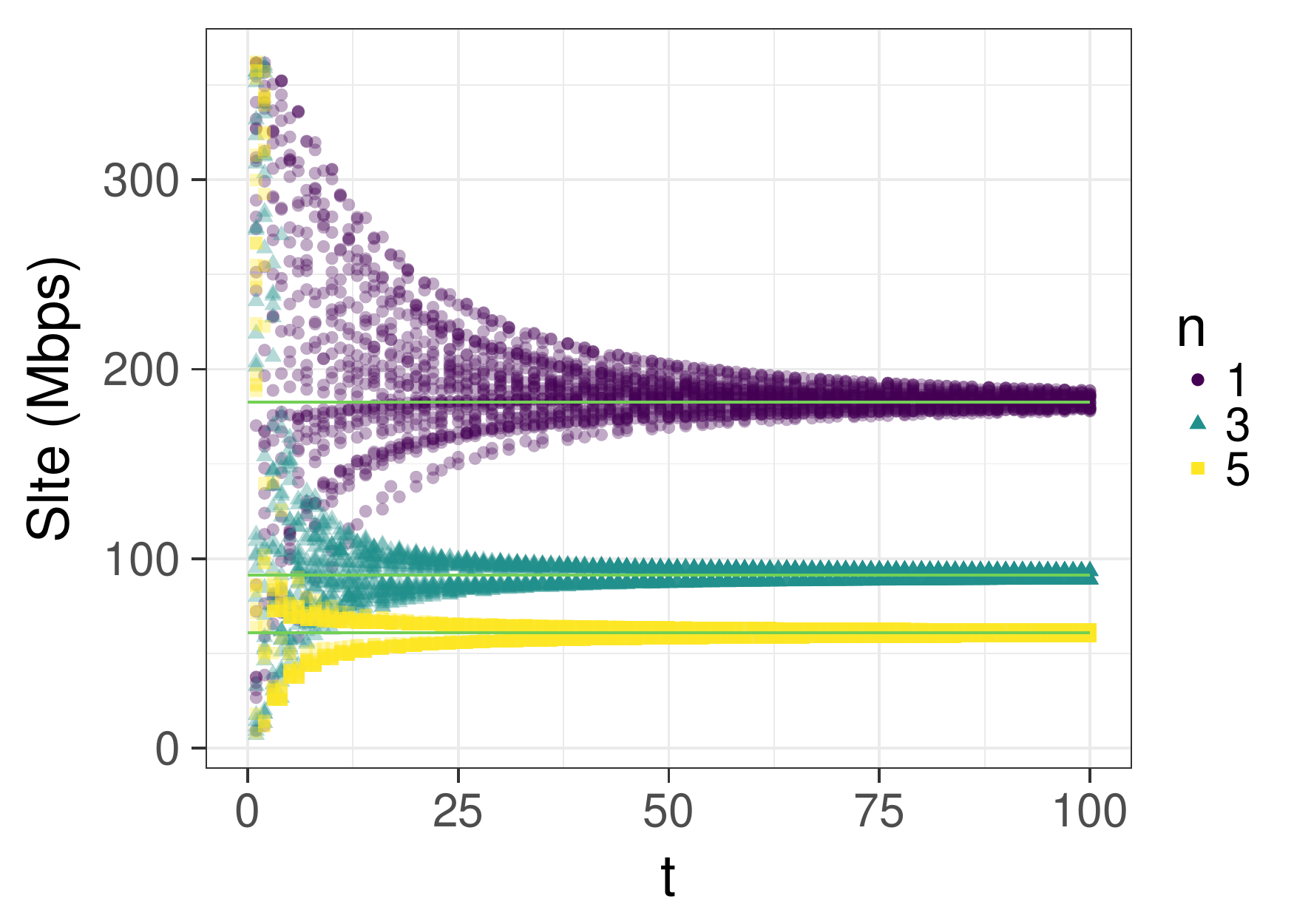}}
\subfigure[$\omega = 1$.]{\includegraphics[width=0.65\columnwidth]{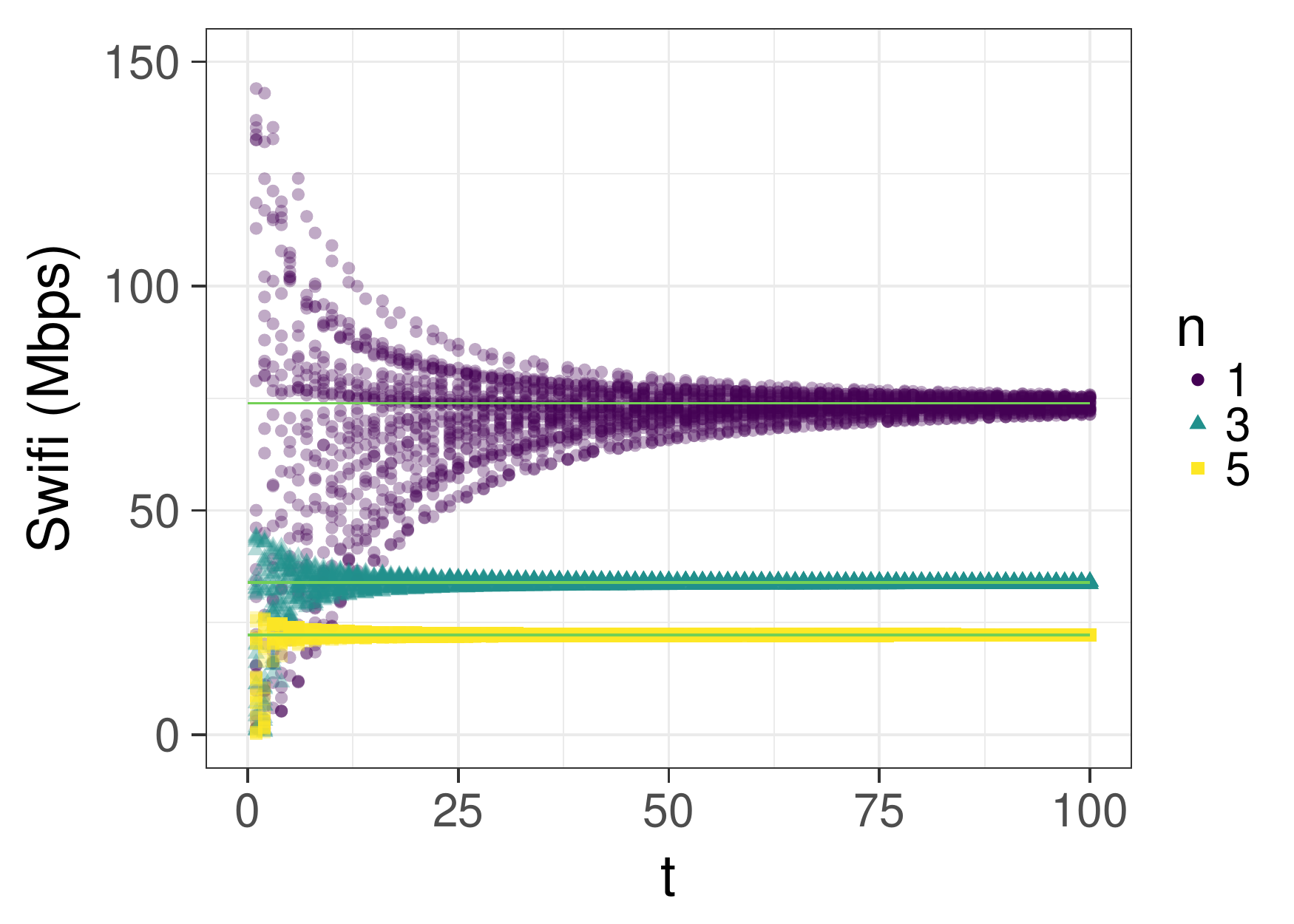}}
\caption{Results varying input parameter $\omega$ and using update rule $\delta_k=\alpha_k = \omega / k^{3/4}$ for $25$ simulation runs. Optimal results depicted as straight lines.}
\label{fig:fixed_n_varying_c1}
\end{figure*}

\section{Performance Evaluation}\label{sec:results}
In this section we evaluate different aspects of the performance of our algorithm \ogdsemp presented in Section~\ref{sec:seq_bco} when 
applied to the unlicensed LTE/WiFi use case formulated in the previous section. For all experiments we have set the gradient-descent step 
size as $\eta_k = 1/k^{1/2}$ and $\delta_k=\alpha_k = \omega / h(k)$, with $\omega$ as input parameter and $h(k)$ as some increasing 
function. We will refer to $\omega$ as the \emph{exploration parameter} and $h$ as the \emph{exploration schedule}. We first evaluate the 
performance of the algorithm, in terms of time of convergence and resulting throughput, while exploring its sensitivity to the input 
parameter $\omega$ and different update rules. We also evaluate the ability of the algorithm to handle network dynamics and quantify its 
performance when the feedback are noisy estimates of the cost function coming from an LTE/WiFi packet network simulator. 

Before we present the results, we provide some high-level comments on the algorithm. We first remark that our initial experiments 
used the algorithm of \citet{FKM05}, although we quickly discarded this idea as we observed painstakingly slow convergence for all ranges 
of parameters. Second, observe that our choice of parameters for \ogdsemp is only supported by theory for the case of infrequently changing 
loss functions, and even then, some tuning was necessary to find the range of constants that lead to good performance. This is a common 
issue with theoretically motivated algorithms: the parameter choices suggested by theory are typically too conservative for practical use. 
Therefore, our experiments can be seen to validate the general algorithm-design principle rather than the exact implementation 
suggested by theory.

\subsection{Sensitivity to the Exploration Parameters}
We first evaluate the performance of the algorithm while varying the exploration parameter $\omega$, which controls how far from $x_t$ we 
take the two cost function evaluations at consecutive iterations.
Note that $\omega$ is then scaled by $h(k)$, which is considered in this section equal to $k^{3/4}$. 
Our interest is in having low variability of results when varying $\omega$ as it is a parameter that depends on the use case and it may be 
hard to optimise in practice. 

Fig.~\ref{fig:fixed_n_varying_c1} shows the resulting $\bar{T}_{{\rm off}}$, ${s}_{{\rm LTE}}$ and $\sum_{j=1}^{n} s_{{\rm wifi},j}/n$ at 
each iteration for $25$ simulation runs and for different values of $n$ (kept fixed during the simulation duration). 
Results are obtained from a Matlab implementation of the algorithm with cost function evaluations computed using Eq.~\eqref{eq:f}, IEEE 
802.11ac parameters as in \cite{ccano-ton} and $T_{\rm on} = 50$ ms. 
We consider the number of packets (of size $1500$ bytes) aggregated in each WiFi transmission equal to $5$.
We have also set $\mathcal K = [-6.9, 0]$ and we evaluate $w = \{0.01, 0.1, 1\}$. 
Note that we have considered $\omega$ to be different orders of magnitude smaller than the diameter of the set $\mathcal K$.
Optimal results from \cite{ccano-ton} are also depicted in Fig.~\ref{fig:fixed_n_varying_c1} as straight lines.

First, we observe in Fig.~\ref{fig:fixed_n_varying_c1} that, for the range of values considered, parameter $\omega$ has slight impact on the rate of convergence. 
However, we see that with $\omega=1$ and especially for $n=5$ results are concentrated in two trajectories. 
The cause for this are notable oscillatory effects which persist even for $t>25$.
Nevertheless, the required number of iterations for convergence is low for all cases. 
That is, in less than $50$ iterations results for $\bar{T}_{{\rm off}}$ are at most $20$ ms apart from the optimal ones. 
In terms of throughput, we see that at $t=50$ resulting throughput is at most $25$ Mbps (for LTE) and $12$ Mbps (for WiFi) off from the 
optimum.

\begin{figure*}[hhht!] 
\centering
\subfigure[$h(k) = k^{3/4}$, $\omega = 0.01$.]{\includegraphics[width=0.67\columnwidth]{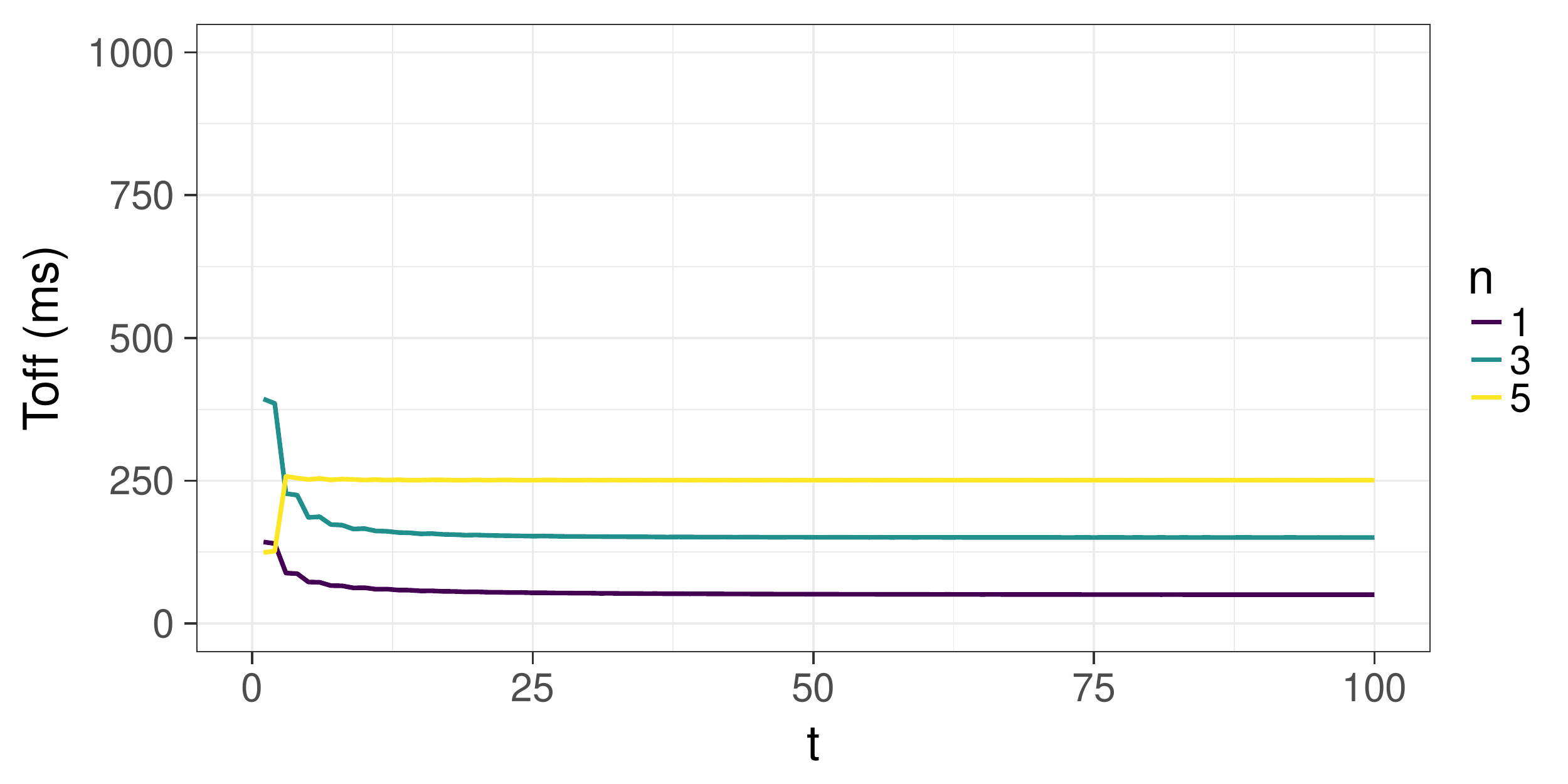}}
\subfigure[$h(k) = k^{3/4}$, $\omega = 0.1$.]{\includegraphics[width=0.67\columnwidth]{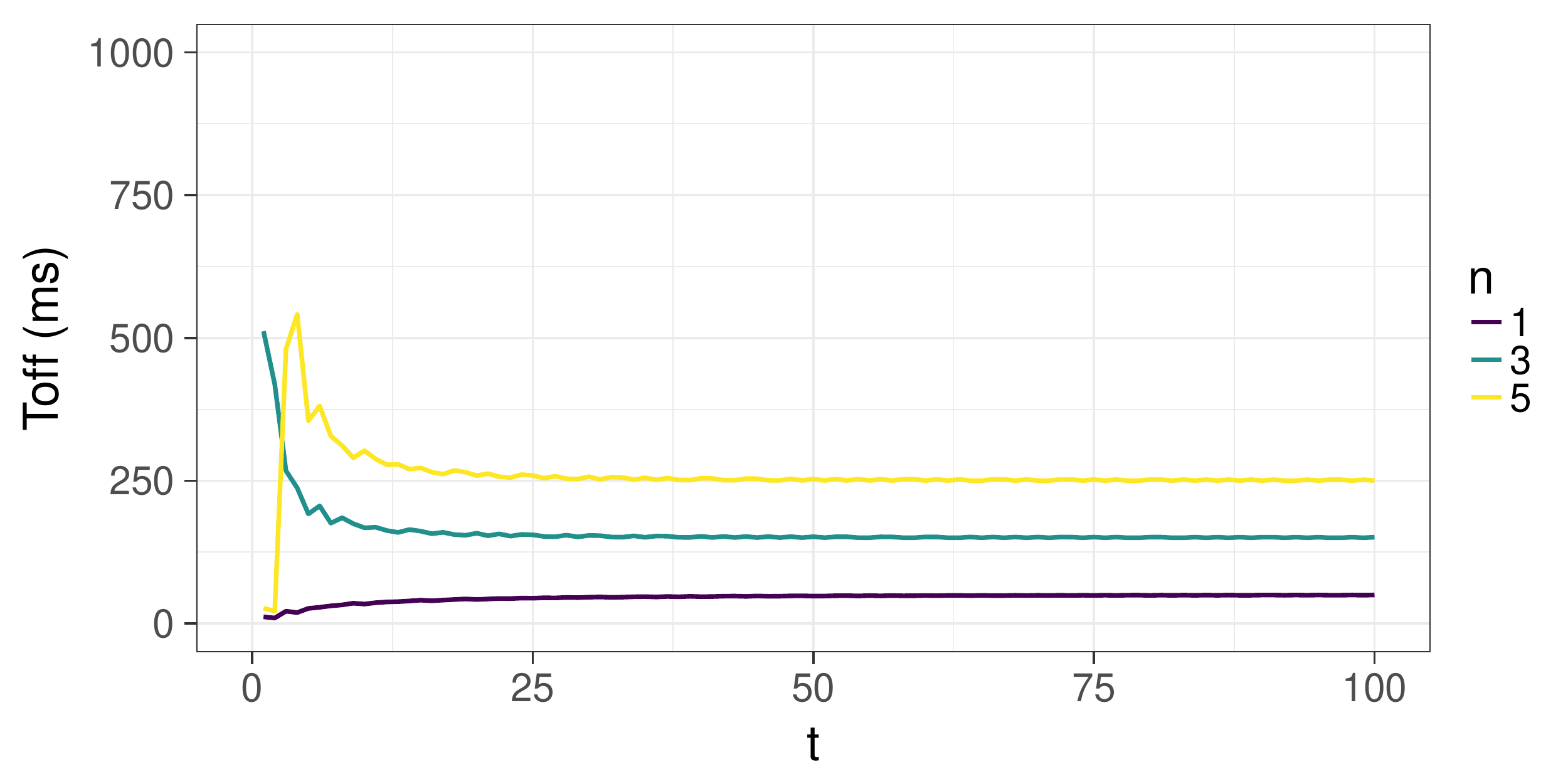}}
\subfigure[$h(k) = k^{3/4}$, $\omega = 1$.]{\includegraphics[width=0.67\columnwidth]{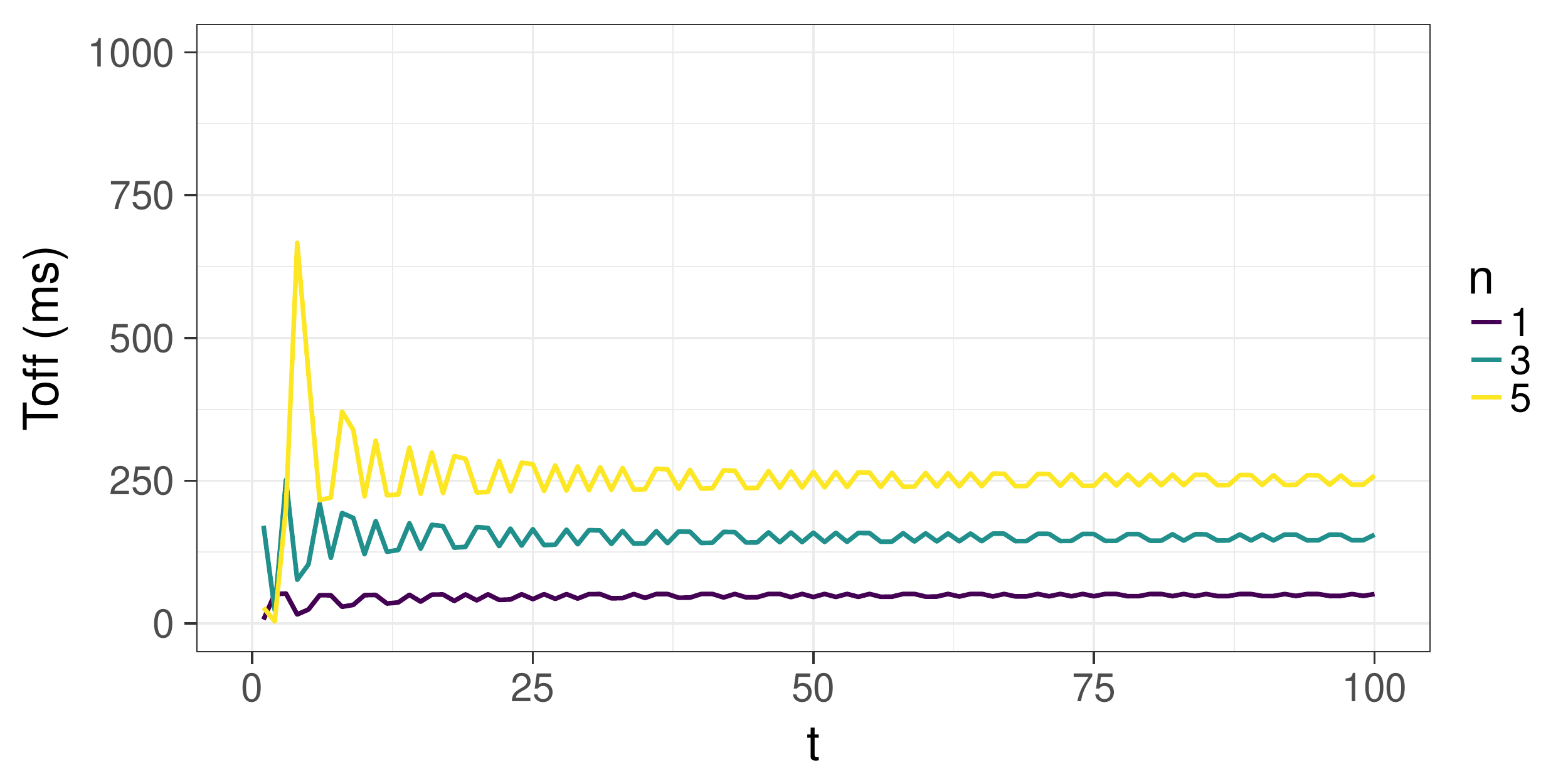}}
\subfigure[$h(k) =  k^{1/2}$, $\omega = 0.01$.]{\includegraphics[width=0.67\columnwidth]{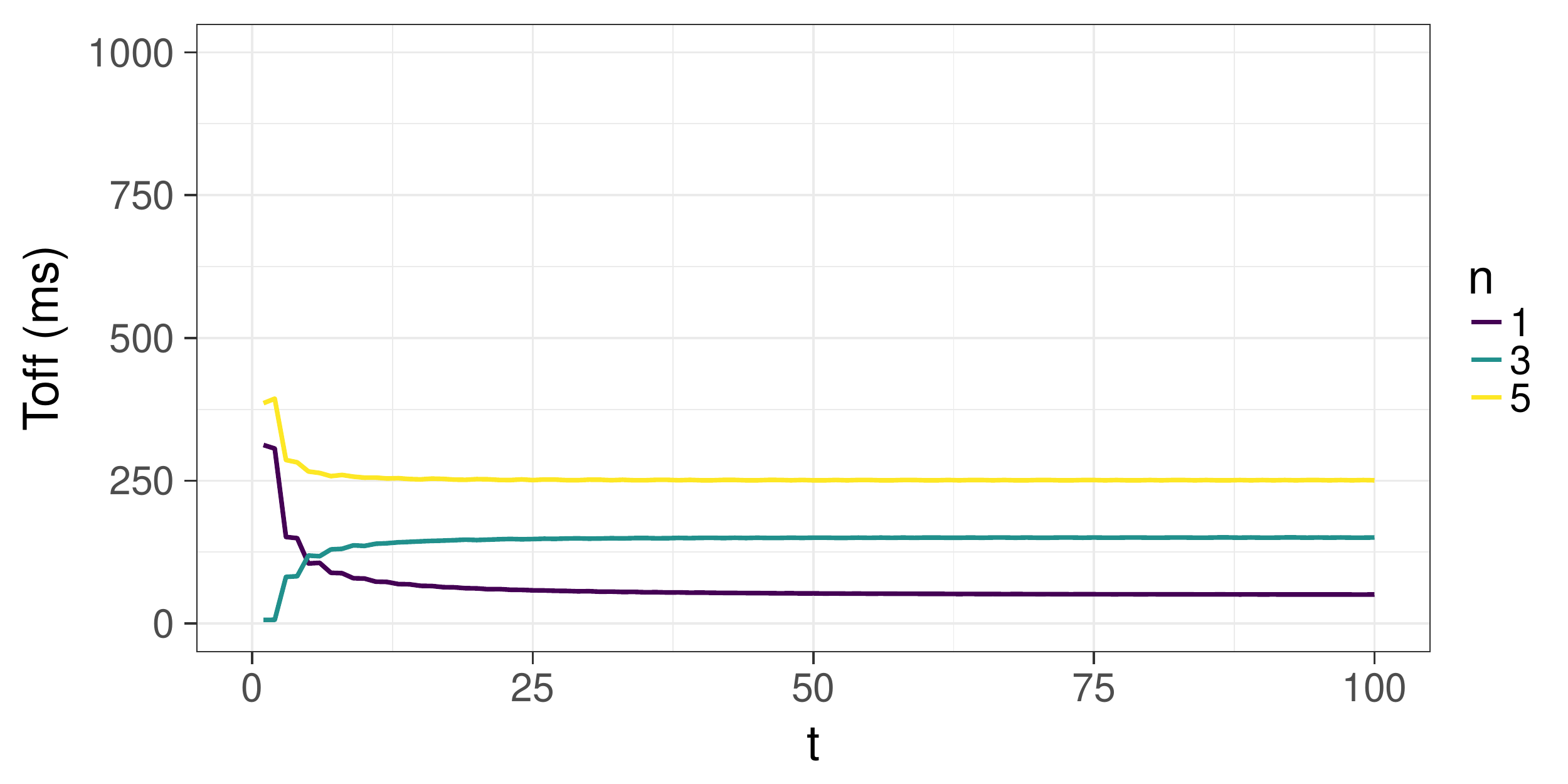}}
\subfigure[$h(k) =  k^{1/2}$, $\omega = 0.1$.]{\includegraphics[width=0.67\columnwidth]{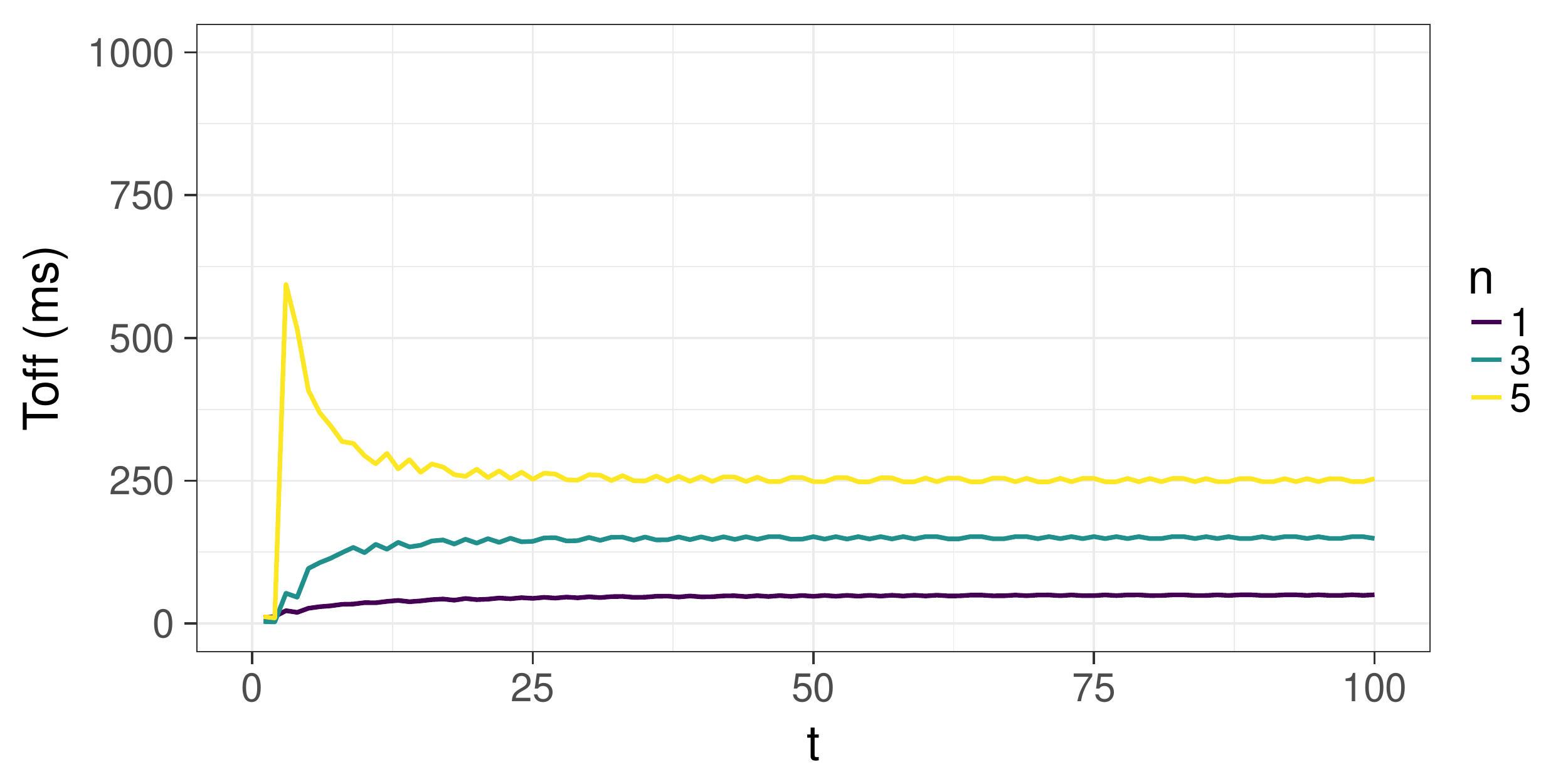}}
\subfigure[$h(k) =  k^{1/2}$, $\omega = 1$.]{\includegraphics[width=0.67\columnwidth]{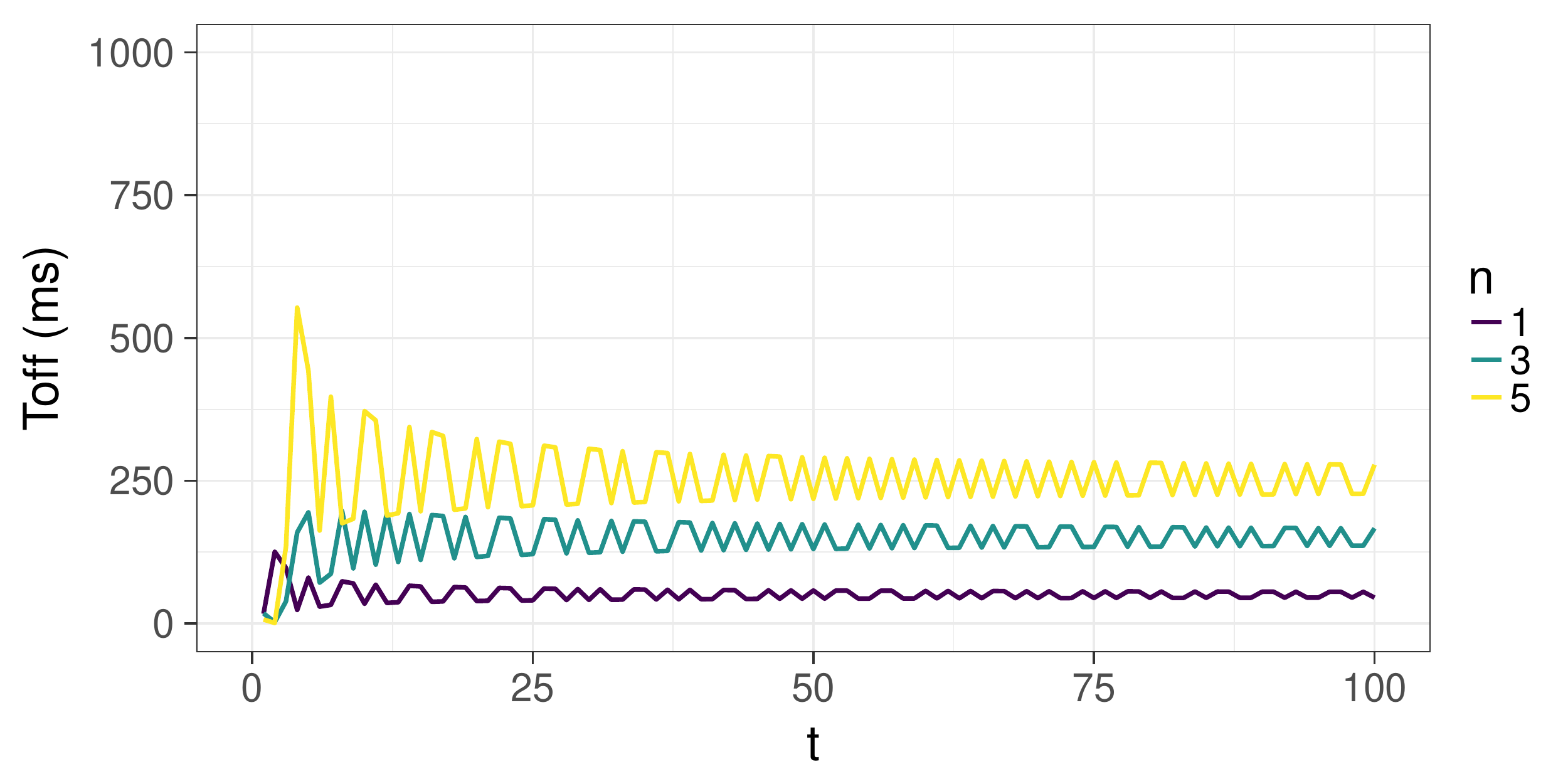}}
\caption{Temporal results using different $\omega$ and update rules $h(k) = k^{3/4}$ and $h(k) = k^{1/2}$ for a single simulation run.}
\label{fig:fixed_n_varying_updaterule}
\end{figure*}

\subsection{Sensitivity to Different Exploration Schedules}

Using the same setup as before, we evaluate now the sensitivity of the algorithm to different exploration schedules. 
Again, our interest is in having low variability of results but noting that this parameter directly controls convergence.

In order to have a clearer picture of the oscillatory effects, we plot now the temporal evolution of $\bar{T}_{{\rm off}}$ for a single 
simulation run in Fig.~\ref{fig:fixed_n_varying_updaterule}, for different values of $\omega$ and exploration schedules $h(k) = k^{3/4}$ and 
$h(k) = k^{1/2}$. 
We can see that although fluctuation due to oscillations increases slightly with the schedule $h(k) = k^{1/2}$, the effect is not 
considerable for $\omega=0.01$ and $\omega=0.1$ (Fig.~\ref{fig:fixed_n_varying_updaterule}a-b and 
Fig.~\ref{fig:fixed_n_varying_updaterule}d-e).
However, when using $\omega=1$ (Fig.~\ref{fig:fixed_n_varying_updaterule}c and Fig.~\ref{fig:fixed_n_varying_updaterule}f), fluctuations of magnitude close to $50$ ms are persistent even at $t=100$ when $h(k) = k^{1/2}$ and $n=5$.
These results suggest that both $\omega$ and $h(k)$ should be selected jointly and among a good-performing range depending on the use case.
However, we expect variations among this range to provide similar results.

\subsection{Adaptability to Network Dynamics}

\begin{figure*}[hhht!] 
\centering
\subfigure[$n$ increases in $1$.]{\includegraphics[width=0.65\columnwidth]{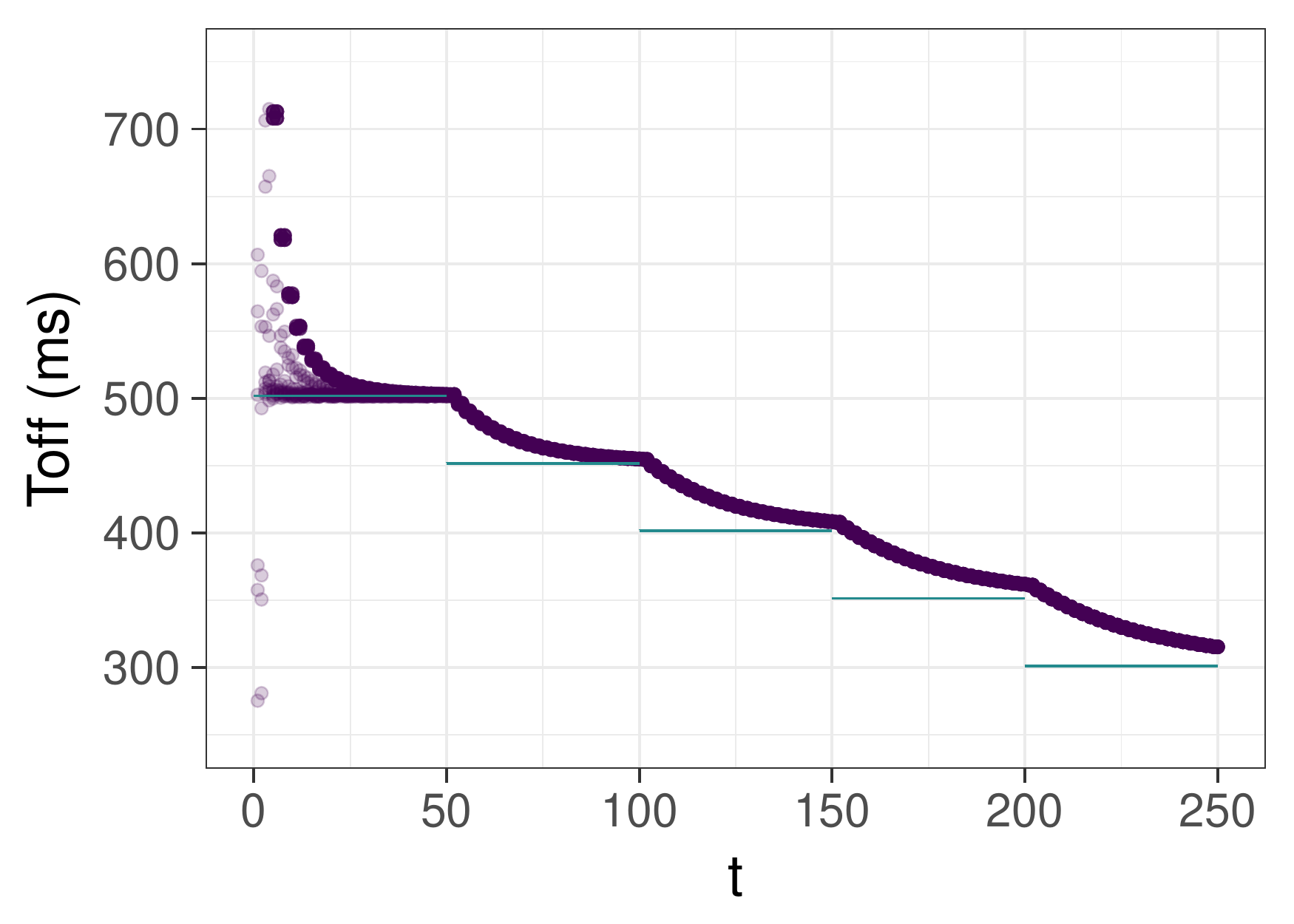}}
\subfigure[$n$ increases in $1$.]{\includegraphics[width=0.65\columnwidth]{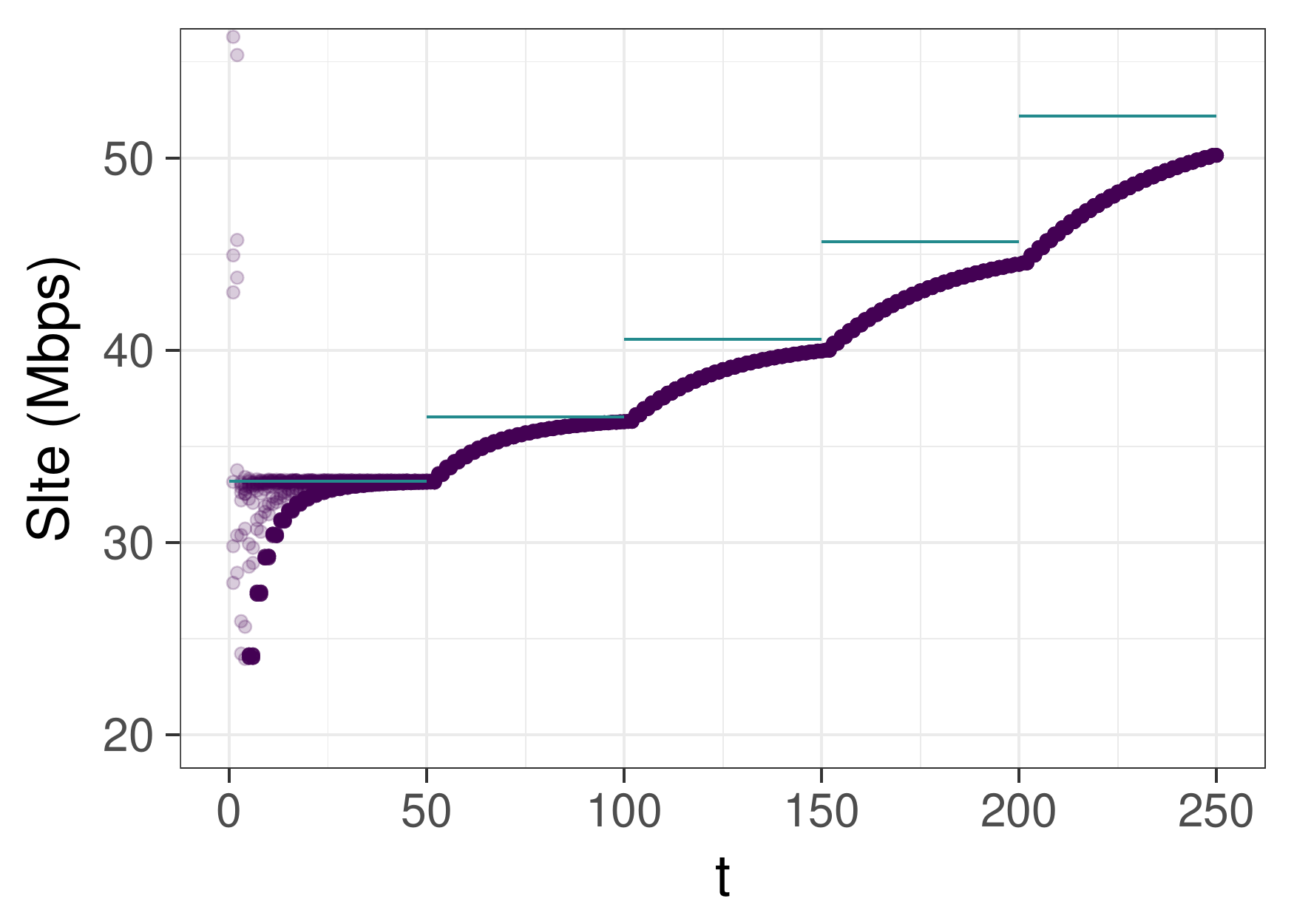}}
\subfigure[$n$ increases in $1$.]{\includegraphics[width=0.65\columnwidth]{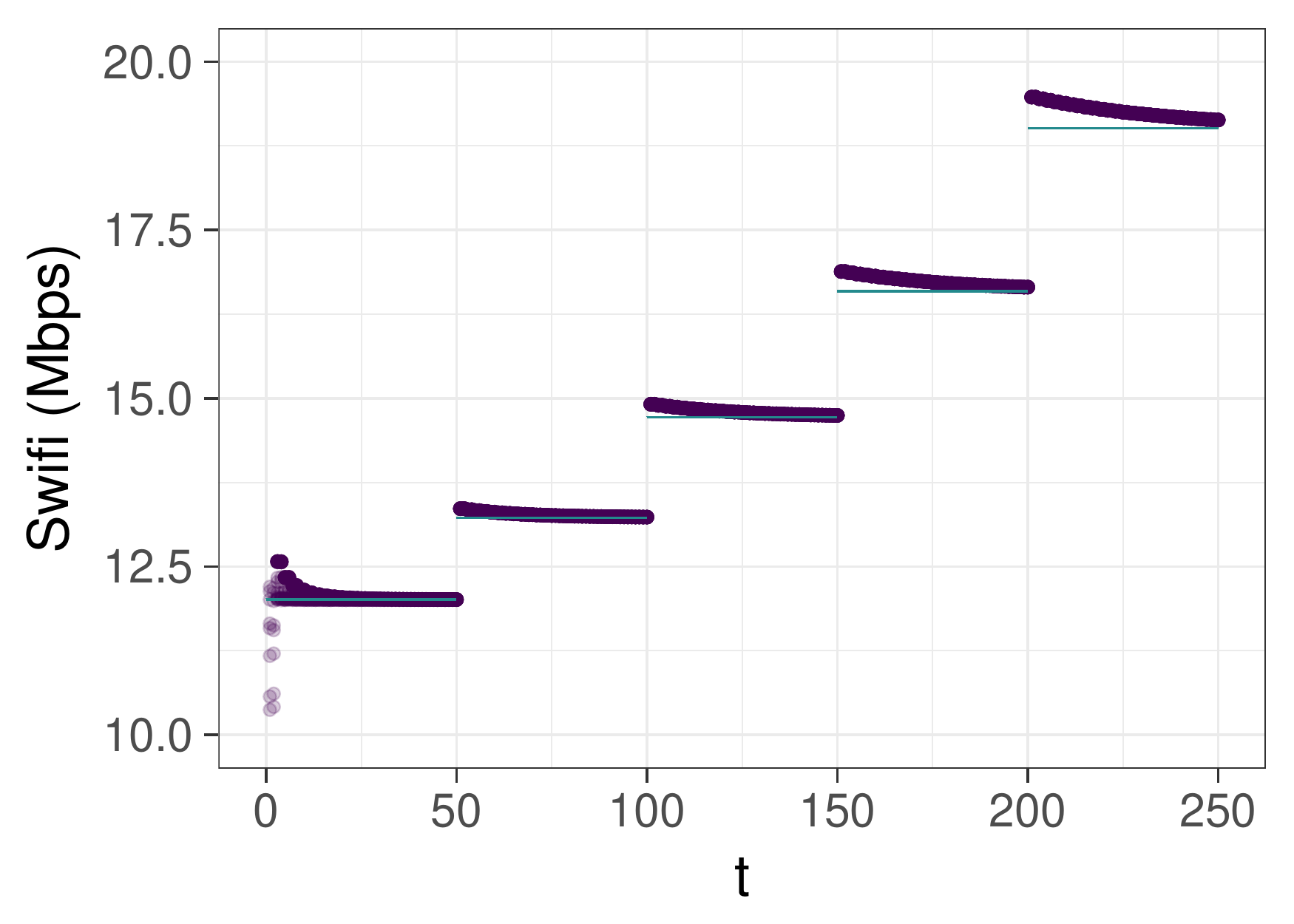}}
\subfigure[$n$ decreases in $1$.]{\includegraphics[width=0.65\columnwidth]{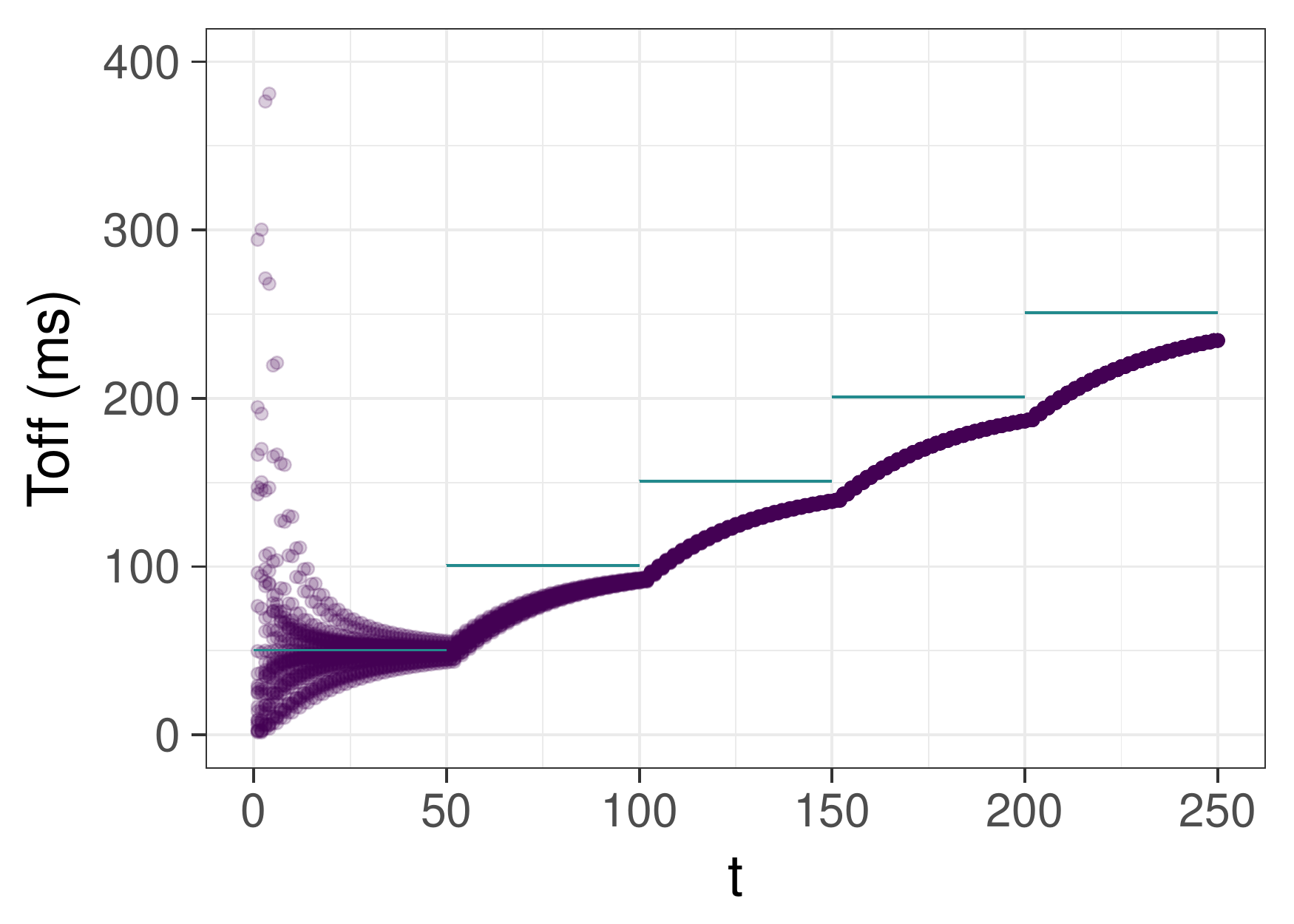}}
\subfigure[$n$ decreases in $1$.]{\includegraphics[width=0.65\columnwidth]{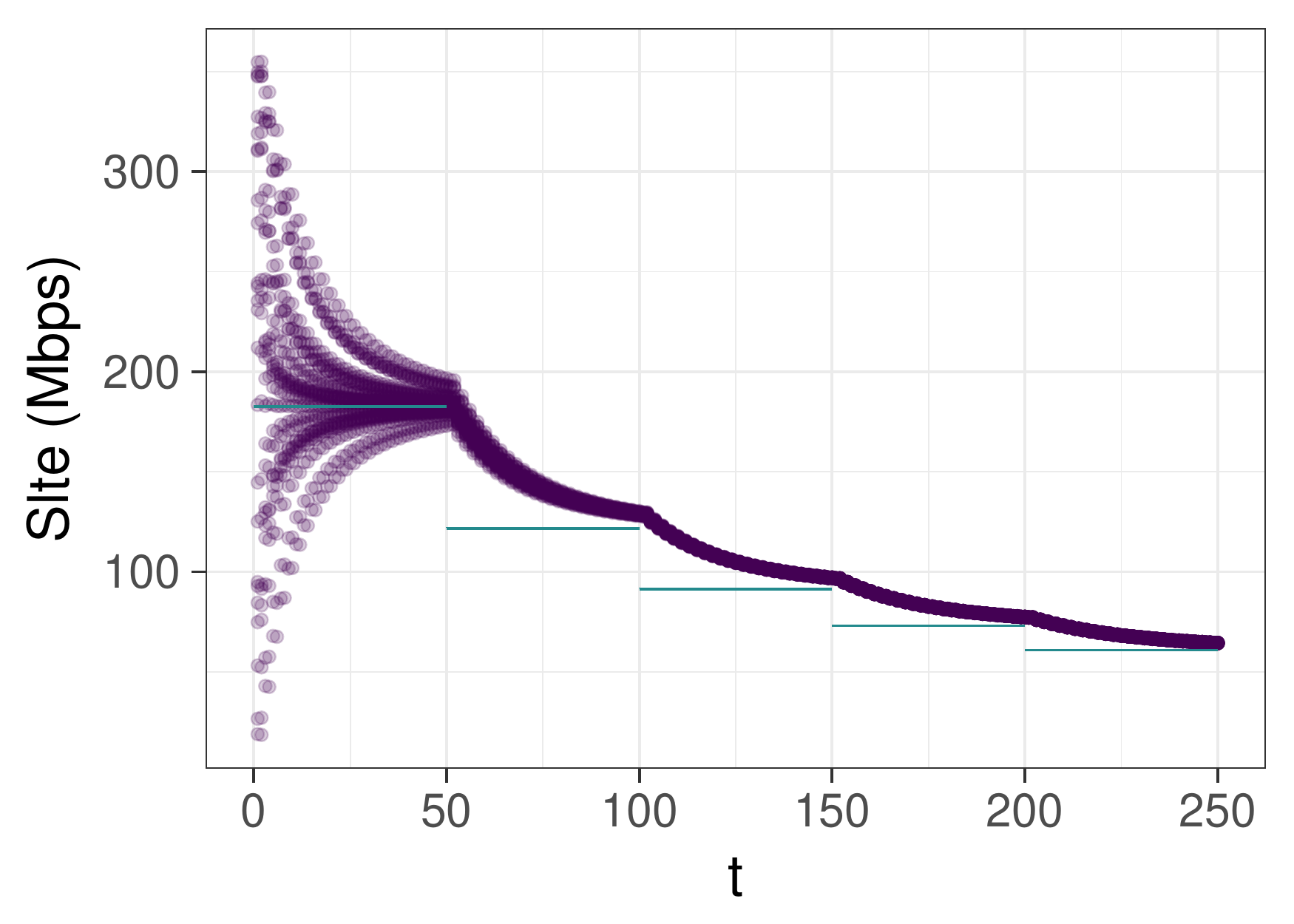}}
\subfigure[$n$ decreases in $1$.]{\includegraphics[width=0.65\columnwidth]{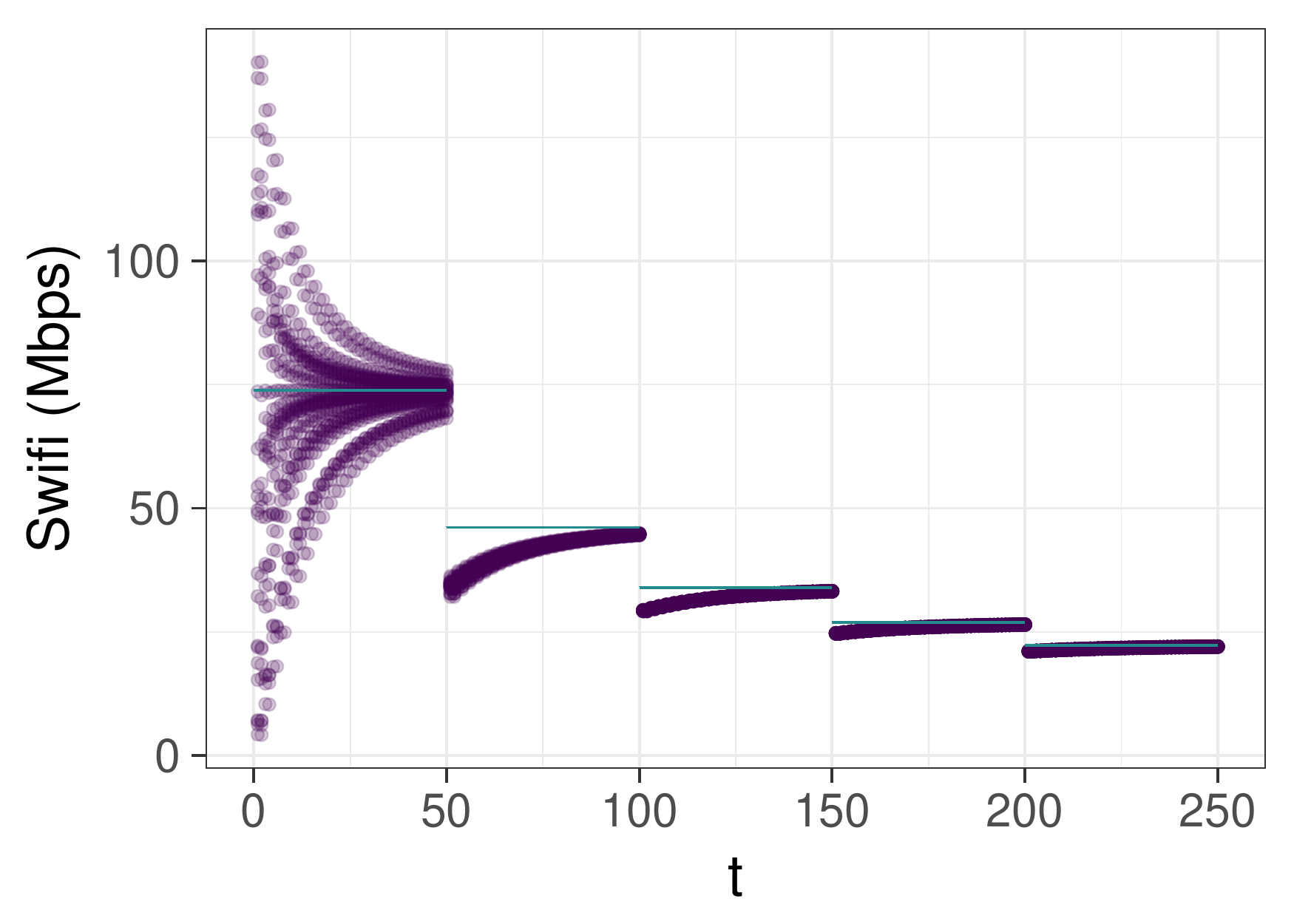}}
\caption{Results changing $n$ ($1$ step increase/decrease each $50$ iterations) with $\omega = 0.01$ and update rule $\delta_k=\alpha_k = \omega / k^{3/4}$ for $25$ simulation runs. Optimal results depicted as straight lines.}
\label{fig:changing_n_slow}
\end{figure*}

\begin{figure}[hhht!] 
\centering
\subfigure[$n=\{10, 5, 10\}$ at $t=\{0, 50, 100\}$.]{\includegraphics[width=0.72\columnwidth]{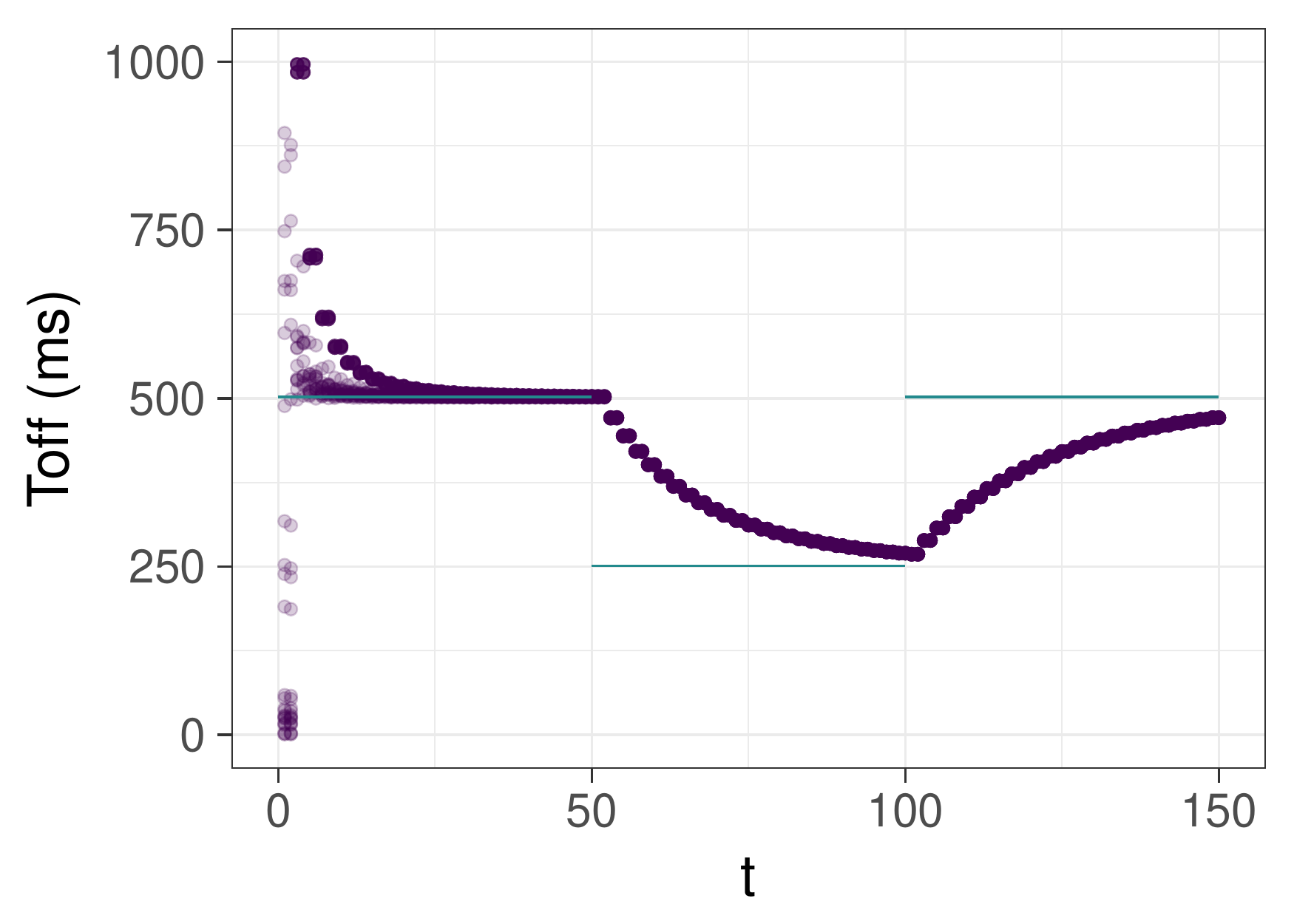}}\\
\subfigure[$n=\{10, 1, 10\}$ at $t=\{0, 100, 200\}$.]{\includegraphics[width=0.72\columnwidth]{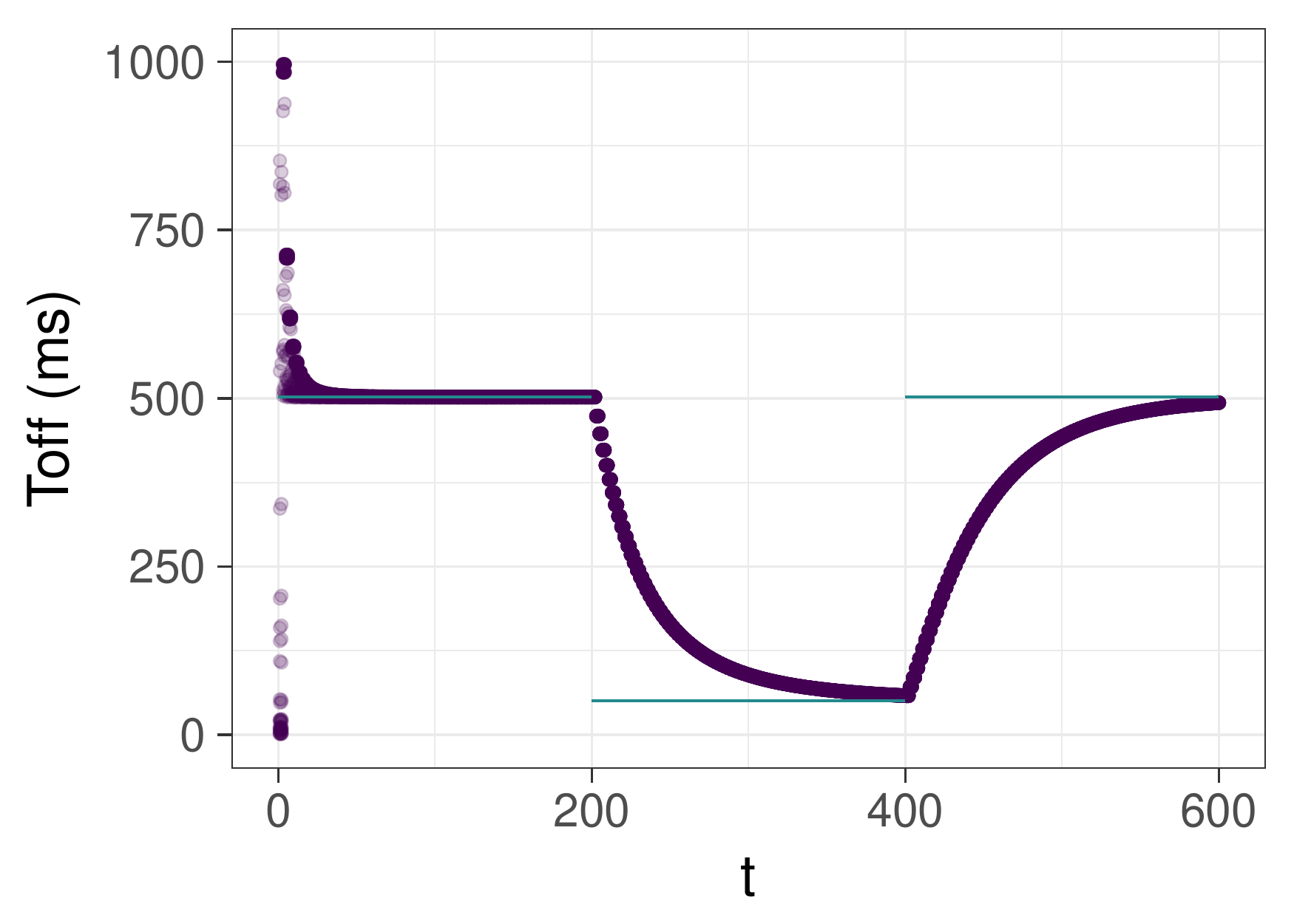}}\\
\caption{Results changing $n$ with $\omega = 0.01$ and update rule $\delta_k=\alpha_k = \omega / k^{3/4}$ for $25$ simulation runs. Optimal results depicted as straight lines..}
\label{fig:changing_n_fast}
\end{figure}

Using again the same setup as in the sections above, we now evaluate the adaptability of the algorithm to network dynamics. 
In particular, we change the number of WiFi stations ($n$) and we do so exactly at the iteration where the gradient is computed so that the incurred error is the highest.
We apply truncation to deal with high divergences in the gradient and when such a case is detected the algorithm uses the gradient computed last. Fig.~\ref{fig:changing_n_slow} shows results when $n$ increases/decreases by 1 each $50$ iterations. 
Fig.~\ref{fig:changing_n_fast} shows faster dynamics, with $n$ changing suddenly in $5$ and $10$ nodes.

We observe in Fig.~\ref{fig:changing_n_slow} how the algorithm is able to quickly adapt to slow changes in the number of WiFi stations. 
It can be seen that the latest the change takes place the slower the algorithm reacts.
The cause for this is the reduction of both the gradient descent step size ($\eta_k$) and the distance from the sample point $x_t$ ($\delta_k$) with $k$.
Despite this we see that the biggest resulting errors in throughput are no greater than $10\%$.

When we consider faster dynamics and vary $n$ by $5$ from one iteration to the next (Fig.~\ref{fig:changing_n_slow}a), we can observe that the algorithm is still able to converge to the new settings in the same amount of iterations. 
When the number of nodes now changes from $1$ to $10$ and viceversa (Fig.~\ref{fig:changing_n_slow}b), we see how the algorithm takes substantially longer to converge.
It is worth noting that periodic adaptation of the learning parameters could address faster changes if those were to be expected.
Since it is not the case in a wireless network scenario we opt for a fixed setting as it is easier to implement in practice.

\subsection{Sensitivity to Noisy Gradient Estimates}

We now quantify the effects on performance of having noisy estimates of the cost function instead of the true value. 
With this goal, we implement the algorithm in a custom packet network simulator implementing LTE and WiFi channel accesses.
Noise sources in this use case include the randomisation of $\bar{T}_{{\rm off}}$ and slot transmission probabilities by WiFi nodes.

We take as cost function the average WiFi and LTE throughput experienced during each temporal batch (simulation time between iterations of the algorithm). 
We note that in this setup the algorithm becomes more sensitive to the exploration parameter $\omega$, that is, the closer to $x_t$ we 
evaluate the cost function, the bigger the effect of the noise in the estimate of the gradient.
These effects are alleviated by increasing the duration of the temporal batch so that the average of the WiFi and LTE throughput become more accurate. 
In this work, we set the temporal batch equal to $t_{\rm b}=50$ s and $\omega=1$ and leave as future work the optimisation of $t_{\rm b}$ as well as methods to deal with noisy estimates resulting from shortening the temporal batch duration.
These methods can include, for instance, averaging the gradient estimates across multiple samples.

\begin{figure*}[hhht!] 
\centering
\subfigure[]{\includegraphics[width=0.67\columnwidth]{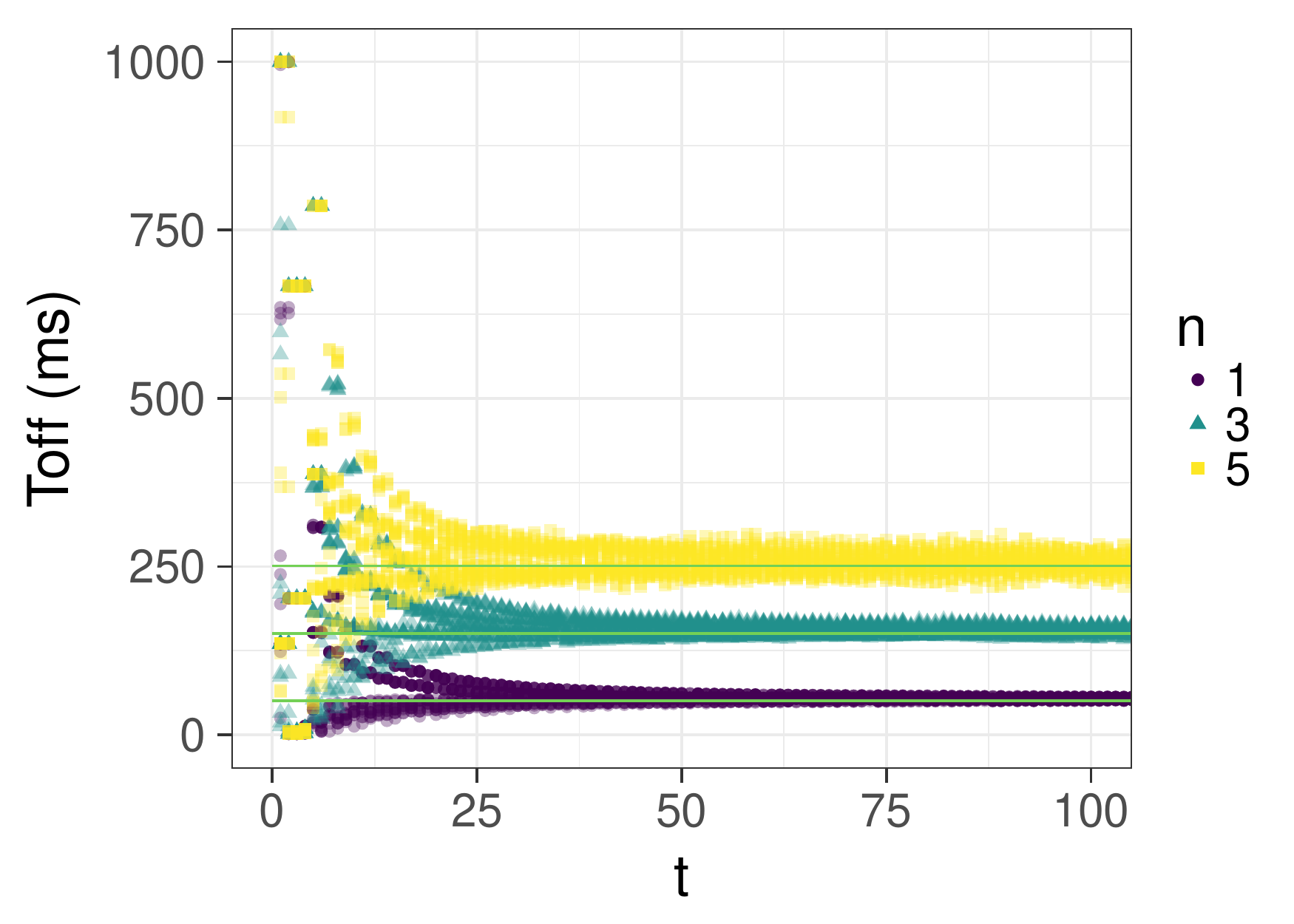}}
\subfigure[]{\includegraphics[width=0.67\columnwidth]{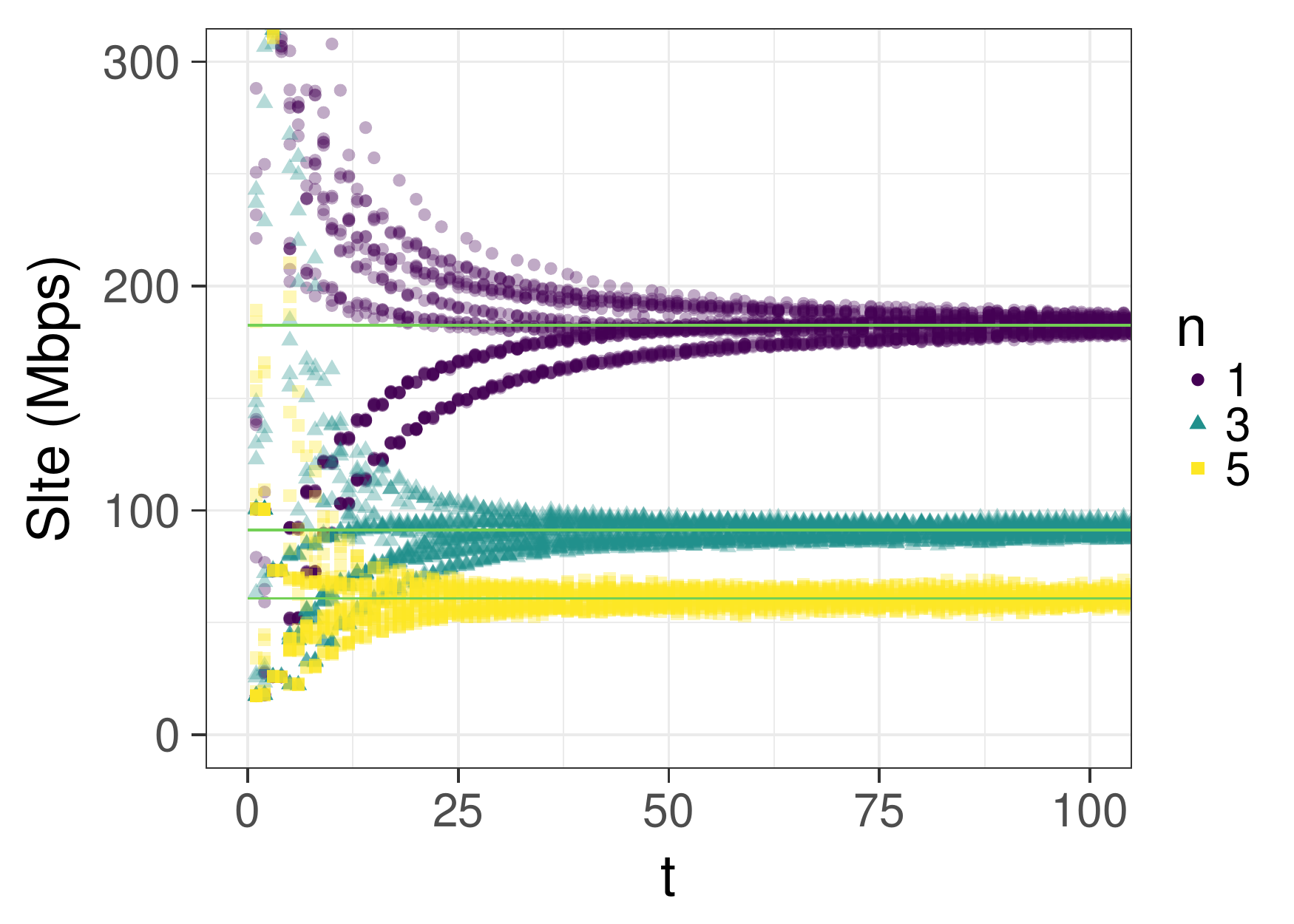}}
\subfigure[]{\includegraphics[width=0.67\columnwidth]{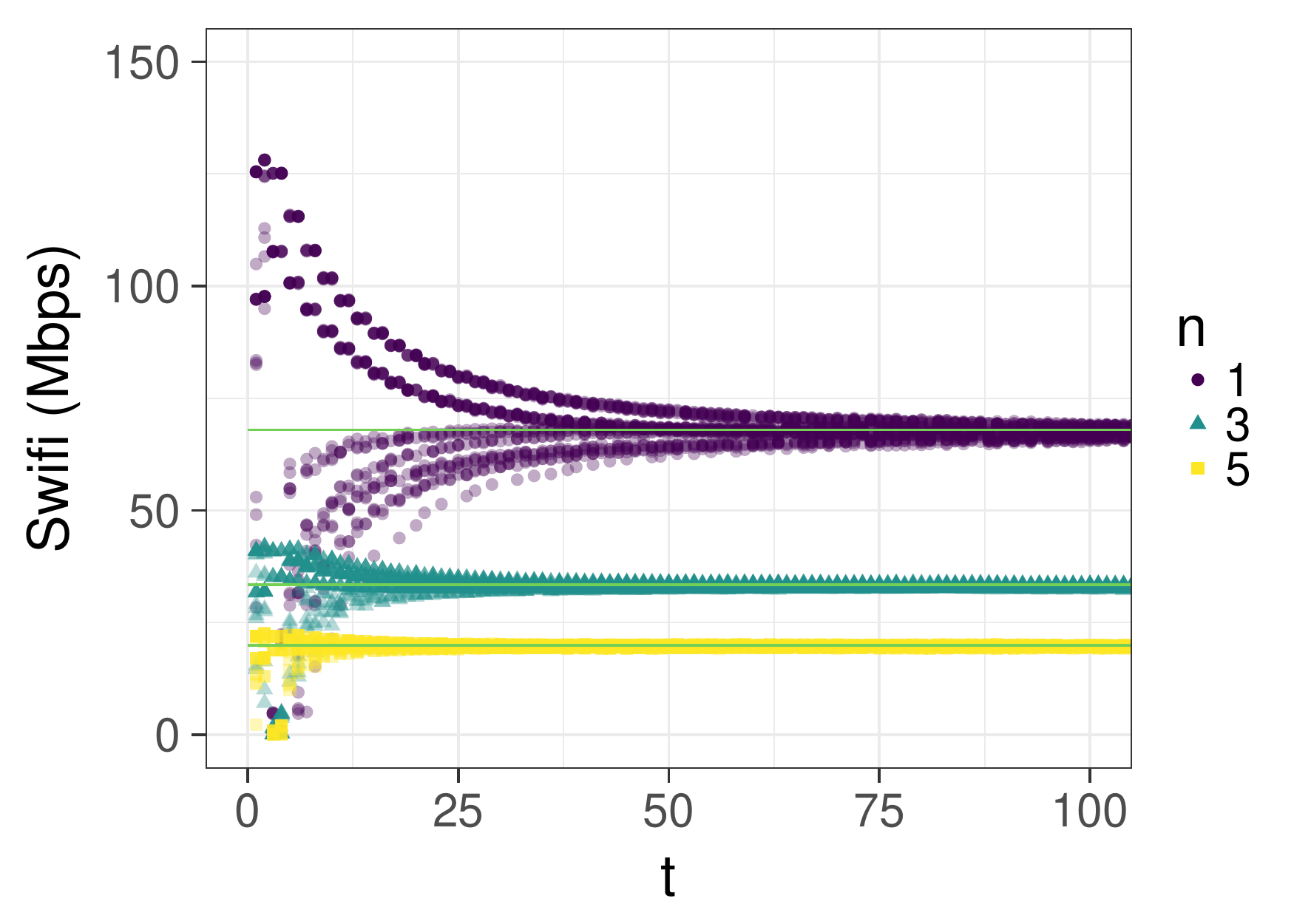}}
\caption{Simulation results with $\omega=1$, update rule $h(k) = k^{3/4}$ and $t_{\rm b}=50$ for $25$ simulation runs. Optimal results depicted as straight lines.}
\label{fig:fixed_n_noisy_estimates}
\end{figure*}

Results for $25$ simulation runs with $\omega=1$ and $h(k) = k^{3/4}$ are shown in Fig.~\ref{fig:fixed_n_noisy_estimates}.
Note that compared to Fig.~\ref{fig:fixed_n_varying_c1}c, we observe higher variability for higher $t$.
This result is to be expected as noisy cost function evaluations can make gradient descent to move uphill.
We observe that the higher $n$ the higher the variability, which is to be expected due to the randomisation of the slot transmission 
probabilities of the WiFi nodes. Despite all this, we can see that for these settings the algorithm still converges in throughput in a 
small number of iterations, that is, at $t=50$ throughput is at most $25$ Mbps (for LTE) and $12$ Mbps (for WiFi) far from the optimal for 
all cases.
Note as well that, despite the $\bar{T}_{{\rm off}}$ variability observed even for large $t$, the effects on throughput are small (at 
$t=100$ the throughput is at most $5$ Mbps off from the optimum for LTE and WiFi).

%% file: conclusions.tex
\section{Final Remarks}\label{sec:conclusions}

\redd{In this article we brought technical tools from Bandit Convex Optimisation to the field of wireless networking optimisation, and 
demonstrated the power of this approach in a challenging use case. We devised and tested a simple and natural algorithm for this setting, 
and verified that our method is suitable to practical implementation: It converges in a small number of iterations, its sensitivity to the 
input parameters and different update rules is low and is able to handle dynamics and noisy estimates satisfactorily.}

\redd{Our results confirm that BCO is a useful framework for addressing important practical aspects of wireless optimisation that 
haven't received much attention before. In particular, most existing works on wireless optimisation make strong assumptions about the 
environment such as stationarity or full observability, whereas our framework natively handles non-stationary environments and 
partial observability. Since these are inherent properties of our approach, our insights generalize well beyond the particular use case we 
studied here. Accordingly, we expect that our work will have a more general impact in wireless 
networking, inspiring many more researchers in the field to make use of the powerful framework Bandit Convex Optimisation.}



%% file: appendix.tex
\section{The proof of Theorem~\ref{thm:main}}\label{app:proof}
The proof largely builds on ideas by \citet{Zin03}, \citet{FKM05} and \citet{agarwal2010optimal}. Our key idea is to define an auxiliary 
BCO problem where consecutive pairs of rounds are grouped together and the loss functions are the average of the original loss functions.
 Precisely, for any odd $t$, we let $k = (t-1)/2$ and define
 \[
  \varphi_k(x) = \frac{f_t(x) + f_{t+1}(x)}{2}
 \]
 in the auxiliary BCO problem. By Assumption~\ref{as:lipschitz}, $f_t$ and $f_{t+1}$ are both $G$-Lipschitz, which also 
implies the $G$-Lipschitzness of $\varphi_k$. Defining the smoothed loss function
 \[
  \wth_k(y) = \EE{\varphi_k(y+\delta\nu)}
 \]
 with $\nu$ distributed uniformly on $[-1,1]$, we observe that
\begin{align*}
 \frac{d}{dy} \EE{\varphi_k(y+\delta \nu)} &= \frac{d}{dy} \int_{-\delta}^{\delta} \frac {1}{2\delta} \cdot \varphi_k(y + v) \, dv 
 \\
 &= \frac{\varphi_k(y + \delta) - \varphi_k(y - \delta)}{2\delta}.
\end{align*}
Furthermore, by the definition of our gradient estimator, we have
\begin{align*}
 \EE{\tg_k} =& \frac{f_t(y_k + \delta) - f_{t+1}(y_k - \delta)}{4\delta} \\
 &+ \frac{f_{t+1}(y_k + \delta) - f_{t}(y_k - \delta)}{4\delta}
 \\
 =& \frac{f_{t}(y_k + \delta) + f_{t+1}(y_k + \delta)}{4\delta} \\
 &- \frac{f_{t}(y_k - \delta) + f_{t+1}(y_k - \delta)}{4\delta}
 \\
 =& \frac{\varphi_k(y_k + \delta) - \varphi_k(y_k - \delta)}{2\delta} = \nabla \wth_k(y_k).
\end{align*}
That is, we have proved that $\tg_k$ is an unbiased estimate of the gradient of the smoothed objective $\wth_k$.
To bound the magnitude of the gradient estimates, let us first consider the case $\varepsilon_k = 1$:
\begin{align*}
\abs{\tg_k} &= \frac{1}{2\delta} \abs{f_t(y_k + \delta) - f_{t+1}(y_k - \delta)}
 \\
 &\le \frac{1}{2\delta} \abs{f_t(y_k + \delta) - f_{t}(y_k - \delta)} + \frac{\alpha_k}{2\delta} \le G + \frac{\alpha_k}{2\delta},
\end{align*}
where the inequalities crucially use the definition of $\alpha_k$ and the Lipschitz property of the losses. 
The complementary case of $\varepsilon_k=-1$ can be handled analogously. For ease of notation in the 
followings, we define $G_k = G + \frac{\alpha_k}{2\delta}$.

We now define the function 
\[
 c_k(x) = \wth_k(x) + x\cdot \pa{\tg_k - \nabla \wth_k(x)}
\]
for all $x\in\K$. It is easy to see that $c_k$ is convex, its gradient satisfies $\nabla c_k(y_t) = \tg_k$, and $\EE{c_k(x)} = \wth_k(x)$ 
holds for all $x$. Following the analysis of \citet{Zin03} (specifically, the proof of his 
Theorem~1), we can show
\[
 \sum_{k=s'}^{r'} \Bpa{c_k(y_k) - c_k(x)} \le \frac{D^2}{\eta} + \frac{\eta}{2} \sum_{k=s'}^{r'} G_k^2
\]
for any $x\in\K_\delta$.
Now, recalling that $\EE{c_k(x_k)} = \wth_k(x_k)$ and $\EE{c_k(x)} = \wth_k(x)$ both hold, we obtain the bound
\[
 \EE{\sum_{k=s'}^{r'} \pa{\wth_k(y_k) - \wth_k(x)}} \le \frac{D^2}{\eta} + \frac{\eta}{2} \sum_{k=s'}^{r'} G_k^2
\]
for any $x\in\K_\delta$.
By the Lipschitzness of $\varphi_k$, we have
\begin{align*}
 \abs{\varphi_k(x) - \wth_k(x)} &= \abs{\varphi_k(x) - \EE{\varphi_k(x+\delta\nu)}}
 \\
 &\le \EE{\abs{\varphi_k(x) - \varphi_k(x+\delta\nu)}} \le G\delta
\end{align*}
for any $x$, where the first inequality is Jensen's. Applying this bound, we obtain
\[
 \EE{\sum_{k=s'}^{r'} \Bpa{\varphi_k(y_k) - \varphi_k(x)}} \le \frac{D^2}{\eta} + \frac{\eta}{2} \sum_{k=s'}^{r'} G_k^2 + 
\delta G \Delta'
\]

It remains to relate this last expression to the real regret. To this end, observe that
\begin{align*}
 \varphi_k(y_k) =& \frac 12 \pa{f_{t}(x_t - \delta \varepsilon_k) + f_{t+1}(x_{t+1} + \delta \varepsilon_k)}
 \\
 \ge& \frac 12 \pa{f_t(x_t) + f_{t+1}(x_{t+1})} - G\delta,
\end{align*}
where the last step follows from the Lipschitz property of the losses.
Combining this inequality with the previous bound, we get
\begin{align*}
 &\EE{\sum_{t=s}^{r} \Bpa{f_t(x_t) - f_t(x)}} \le \frac{2 D^2}{\eta} + \frac{\eta}{2} \sum_{k=s'}^{r'} G_k^2  + 
3 \delta G \Delta
\\
&\qquad\qquad\le \frac{2 D^2}{\eta} + \frac{\eta G^2 \Delta}{2} + \frac{\eta}{4\delta^2} \sum_{k=s'}^{r'} \alpha_k^2  + 
3 \delta G \Delta,
\end{align*}
where the last step uses the inequality $(a+b)^2 \le 2\pa{a^2 + b^2}$. Combining this bound with the fact that, by the Lipschitzness of 
the loss functions, 
\[
\min_{x\in\K_\delta} f_t(x) \le \min_{x\in\K} f_t(x) + G\delta
\]
holds, the proof is concluded.
\qed